\documentclass[pearl,fleqn,review=false]{jfp-epi}
\RequirePackage{amsmath}
\usepackage{amsfonts}
\usepackage{url}
\usepackage{tikz}
\usepackage{caption}
\usepackage{graphicx}
\usepackage{subcaption}
\usepackage{scalerel}

\makeatletter
\@ifundefined{lhs2tex.lhs2tex.sty.read}%
  {\@namedef{lhs2tex.lhs2tex.sty.read}{}%
   \newcommand\SkipToFmtEnd{}%
   \newcommand\EndFmtInput{}%
   \long\def\SkipToFmtEnd#1\EndFmtInput{}%
  }\SkipToFmtEnd

\newcommand\ReadOnlyOnce[1]{\@ifundefined{#1}{\@namedef{#1}{}}\SkipToFmtEnd}
\usepackage{newtxmath}
\usepackage{stmaryrd}
\DeclareFontFamily{OT1}{cmtex}{}
\DeclareFontShape{OT1}{cmtex}{m}{n}
  {<5><6><7><8>cmtex8
   <9>cmtex9
   <10><10.95><12><14.4><17.28><20.74><24.88>cmtex10}{}
\DeclareFontShape{OT1}{cmtex}{m}{it}
  {<-> ssub * cmtt/m/it}{}

\DeclareFontShape{OT1}{cmtt}{bx}{n}
  {<5><6><7><8>cmtt8
   <9>cmbtt9
   <10><10.95><12><14.4><17.28><20.74><24.88>cmbtt10}{}
\DeclareFontShape{OT1}{cmtex}{bx}{n}
  {<-> ssub * cmtt/bx/n}{}

\newcommand{\Conid}[1]{\mathit{#1}}
\newcommand{\Varid}[1]{\mathit{#1}}
\newcommand{\anonymous}{\kern0.06em \vbox{\hrule\@width.5em}}

\usepackage{polytable}

\@ifundefined{mathindent}%
  {\newdimen\mathindent\mathindent\leftmargini}%
  {}%

\def\resethooks{%
  \global\let\SaveRestoreHook\empty
  \global\let\ColumnHook\empty}
\newcommand*{\savecolumns}[1][default]%
  {\g@addto@macro\SaveRestoreHook{\savecolumns[#1]}}
\newcommand*{\restorecolumns}[1][default]%
  {\g@addto@macro\SaveRestoreHook{\restorecolumns[#1]}}
\newcommand*{\aligncolumn}[2]%
  {\g@addto@macro\ColumnHook{\column{#1}{#2}}}

\resethooks

\newcommand{\onelinecommentchars}{\quad-{}- }
\newcommand{\commentbeginchars}{\enskip\{-}
\newcommand{\commentendchars}{-\}\enskip}

\newcommand{\visiblecomments}{%
  \let\onelinecomment=\onelinecommentchars
  \let\commentbegin=\commentbeginchars
  \let\commentend=\commentendchars}

\newcommand{\invisiblecomments}{%
  \let\onelinecomment=\empty
  \let\commentbegin=\empty
  \let\commentend=\empty}

\visiblecomments

\newlength{\blanklineskip}
\newcommand{\hsindent}[1]{\quad}
\let\hspre\empty
\let\hspost\empty

\EndFmtInput
\makeatother
\ReadOnlyOnce{polycode.fmt}%
\makeatletter

\newcommand{\hsnewpar}[1]%
  {{\parskip=0pt\parindent=0pt\par\vskip #1\noindent}}

\newcommand{\hscodestyle}{}

\newcommand{\sethscode}[1]%
  {\expandafter\let\expandafter\hscode\csname #1\endcsname
   \expandafter\let\expandafter\endhscode\csname end#1\endcsname}

  {\par\noindent
   \advance\leftskip\mathindent
   \hscodestyle
   \let\\=\@normalcr
   \let\hspre\(\let\hspost\)%
   \pboxed}%
  {\endpboxed\)%
   \par\noindent
   \ignorespacesafterend}

  {\hsnewpar\abovedisplayskip
   \advance\leftskip\mathindent
   \hscodestyle
   \let\hspre\(\let\hspost\)%
   \pboxed}%
  {\endpboxed%
   \hsnewpar\belowdisplayskip
   \ignorespacesafterend}

  {\hsnewpar\abovedisplayskip
   \advance\leftskip\mathindent
   \hscodestyle
   \let\\=\@normalcr
   \(\pboxed}%
  {\endpboxed\)%
   \hsnewpar\belowdisplayskip
   \ignorespacesafterend}

\newcommand{\plainhs}{\sethscode{plainhscode}}

\plainhs

  {\hsnewpar\abovedisplayskip
   \advance\leftskip\mathindent
   \hscodestyle
   \let\\=\@normalcr
   \(\parray}%
  {\endparray\)%
   \hsnewpar\belowdisplayskip
   \ignorespacesafterend}

  {\parray}{\endparray}

  {\(\parray}{\endparray\)}

\def\codeframewidth{\arrayrulewidth}
\RequirePackage{calc}

  {\parskip=\abovedisplayskip\par\noindent
   \hscodestyle
   \arrayrulewidth=\codeframewidth
   \tabular{@{}|p{\linewidth-2\arraycolsep-2\arrayrulewidth-2pt}|@{}}%
   \hline\framedhslinecorrect\\{-1.5ex}%
   \let\endoflinesave=\\
   \let\\=\@normalcr
   \(\pboxed}%
  {\endpboxed\)%
   \framedhslinecorrect\endoflinesave{.5ex}\hline
   \endtabular
   \parskip=\belowdisplayskip\par\noindent
   \ignorespacesafterend}

\newcommand{\framedhslinecorrect}[2]%
  {#1[#2]}

  {\(\def\column##1##2{}%
   \let\>\undefined\let\<\undefined\let\\\undefined
   \newcommand\>[1][]{}\newcommand\<[1][]{}\newcommand\\[1][]{}%
   \def\fromto##1##2##3{##3}%
   }{\) }%

  {\let\orighscode=\hscode
   \let\origendhscode=\endhscode
   \def\endhscode{\def\hscode{\endgroup\def\@currenvir{hscode}\\}\begingroup}
   \orighscode\def\hscode{\endgroup\def\@currenvir{hscode}}}%
  {\origendhscode
   \global\let\hscode=\orighscode
   \global\let\endhscode=\origendhscode}%

\makeatother
\EndFmtInput

\newcommand{\circo}{\mathrel{\kern 0.12em%
      \raisebox{1pt}{\tikz \draw[line width=0.6pt] circle(1.1pt);}%
      \kern 0.12em}}

\newlength{\mylen}
\setbox1=\hbox{$\bullet$}\setbox2=\hbox{\tiny$\bullet$}
\newcommand{\myapply}{\mathrel{\kern 0.12em\scalebox{0.8}{\$}\kern 0.12em}}

\renewcommand{\Conid}[1]{{\mathsf{#1}}}

\usepackage{doubleequals}

\newtheorem*{sec-duality}{Second duality theorem}
\newtheorem*{thi-duality}{Third duality theorem}

\def\commentbegin{\quad$\{$~}
\def\commentend{$\}$}
\mathindent=\parindent
\allowdisplaybreaks

\counterwithout{equation}{section}

\begin{document}

\title[Folding the dragon]{Folding the dragon}

\author{Ting-Wu Chang}
\orcid{0009-0007-3860-3738}
\affiliation{%
\institution{Graduate Institute of Networking and Multimedia, National Taiwan University}
\city{Taipei}
\country{Taiwan}
}
\author{Liang-Ting Chen}
\orcid{0000-0002-3250-1331}
\author{Shin-Cheng Mu}
\orcid{0000-0002-4755-601X}
\affiliation{%
\institution{Institute of Information Science,
Academia Sinica}
\city{Taipei}
\country{Taiwan}
}

\begin{abstract}
The Heighway Dragon Curve is one of the best known fractal curves.
There are two ways to construct the curve:
starting from a straight line segment,
repeatedly make a copy of the current curve, rotate it by 90 degrees, and connect
them;
or repeatedly replace each straight segment in the curve by two segments with a right angle.
A natural question is: how do we prove the equivalence of the two approaches?

We generalise the construction of the curve to allow rotations to both sides.
It then turns out that the two approaches are respectively a \ensuremath{\Varid{foldr}} and a \ensuremath{\Varid{foldl}},
and that the key property for proving their equivalence, using the second duality theorem,
is the distributivity of an `interleave' operator.
\end{abstract}

\maketitle

\section{Introduction}

\begin{figure}
\begin{subfigure}[b]{0.4\linewidth}
\centering
\def\svgwidth{4.5cm}
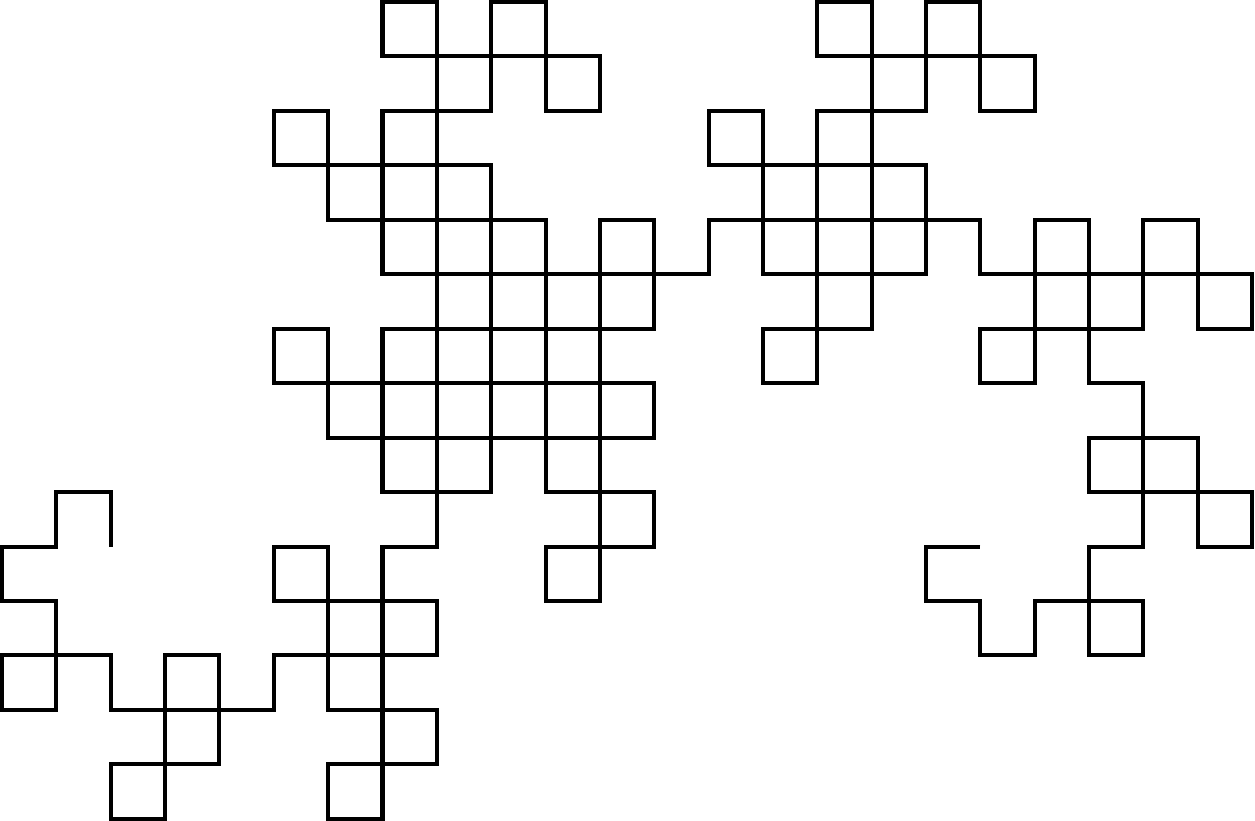
\caption{}
\label{subfig:dragon-8}
\end{subfigure}%
\begin{subfigure}[b]{0.6\linewidth}
\centering
\def\svgwidth{7.4cm}
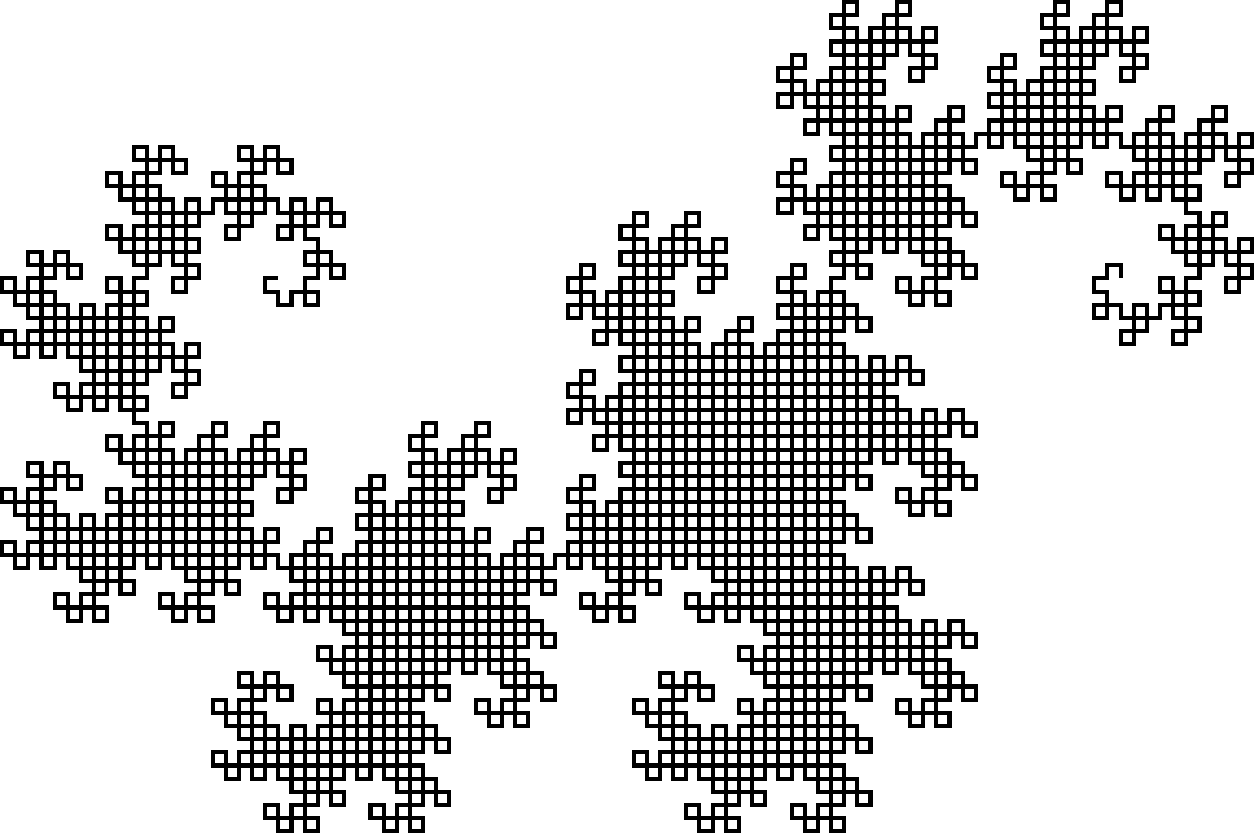
\caption{}
\label{subfig:dragon-12}
\end{subfigure}
\caption{Dragon curves of orders 8 and 12.}
\label{fig:dragon-8-12}
\end{figure}

\begin{figure}
\centering
\includegraphics[width=0.8\textwidth]{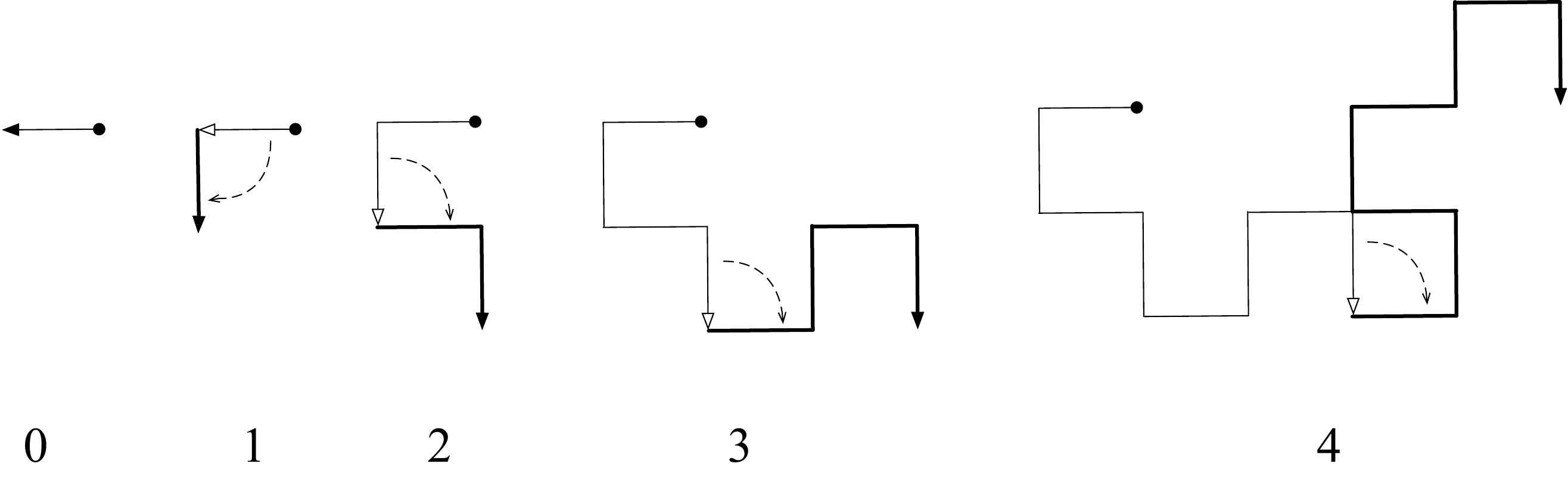}
\caption{The mirroring construction of dragon curves of orders 0 to 4.
The rotated copy is shown in thick lines, and the old endpoint (the center of rotation)
is shown as a hollow-headed arrow.}
\label{fig:dragon-unfold}
\end{figure}

\begin{figure}
\centering
\includegraphics[width=0.8\textwidth]{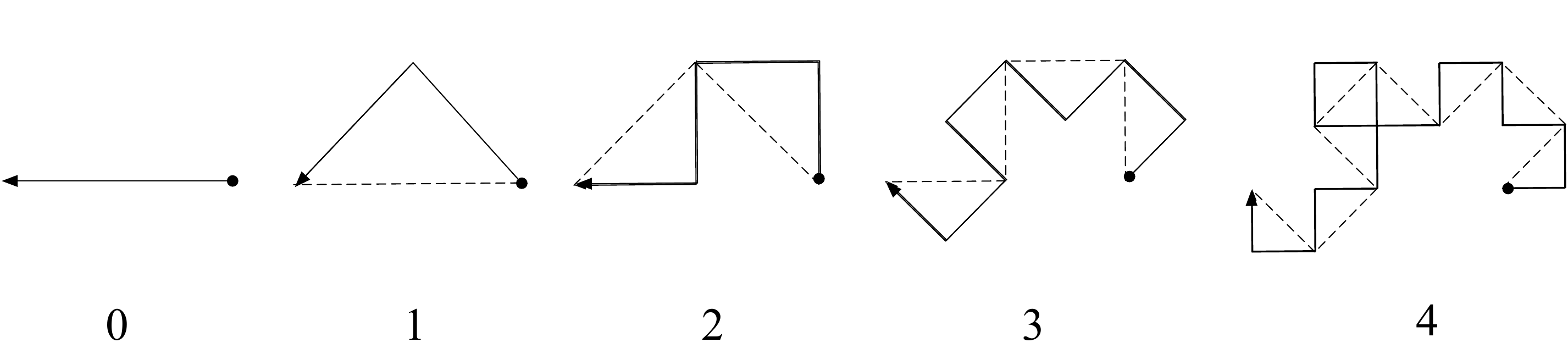}
\caption{The interleaving construction of dragon curves of orders 0 to 4.
The previous curve for each order is shown using dashed lines.}
\label{fig:dragon-interleave}
\end{figure}

The \emph{Heighway Dragon Curve}, discovered by John E. Heighway in 1966, is among the best known recursive fractal curves.
Figures~\ref{fig:dragon-8-12}(a) and~\ref{fig:dragon-8-12}(b) show two examples of dragon curves, of orders \ensuremath{\mathrm{8}} and \ensuremath{\mathrm{12}}, respectively ---
the concept of `order' will soon be discussed.
There are two ways to generate the curves.
In what we will refer to as the {\bf mirroring} approach,
we start with a single line --- which is an order \ensuremath{\mathrm{0}} dragon curve,
with its two ends respectively marked as the starting point (a dot) and the end point (an arrow), as shown on the left of Figure~\ref{fig:dragon-unfold}.
To generate the curve of the next order,
we duplicate the current curve, rotate the copy around the end point 90 degrees clockwise,
and let the starting point of the copy be the new end point.
Figure~\ref{fig:dragon-unfold} shows dragon curves of orders \ensuremath{\mathrm{0}} to \ensuremath{\mathrm{4}} generated in this manner.
In another {\bf interleaving} approach
(the rationale behind the naming will become clear later),
we start with the same single line as the order \ensuremath{\mathrm{0}} dragon curve.
To generate the curve of the next order, we traverse through the current curve from the beginning to the end,
while adding a crease on each line segment we encounter.
The creases should turn to the left and the right (of the direction we are going) alternately,
as we can see in Figure~\ref{fig:dragon-interleave}:
the curve of order \ensuremath{\mathrm{1}} is generated by breaking the straight line into a left turn (that is, leaning to the right, then turning left),
the order \ensuremath{\mathrm{2}} curve is generated from that of order \ensuremath{\mathrm{1}} by breaking the first line into a left turn, and the second line to a right turn.

\begin{figure}
\begin{subfigure}[b]{0.68\textwidth}
\centering
\includegraphics[width=0.8\textwidth]{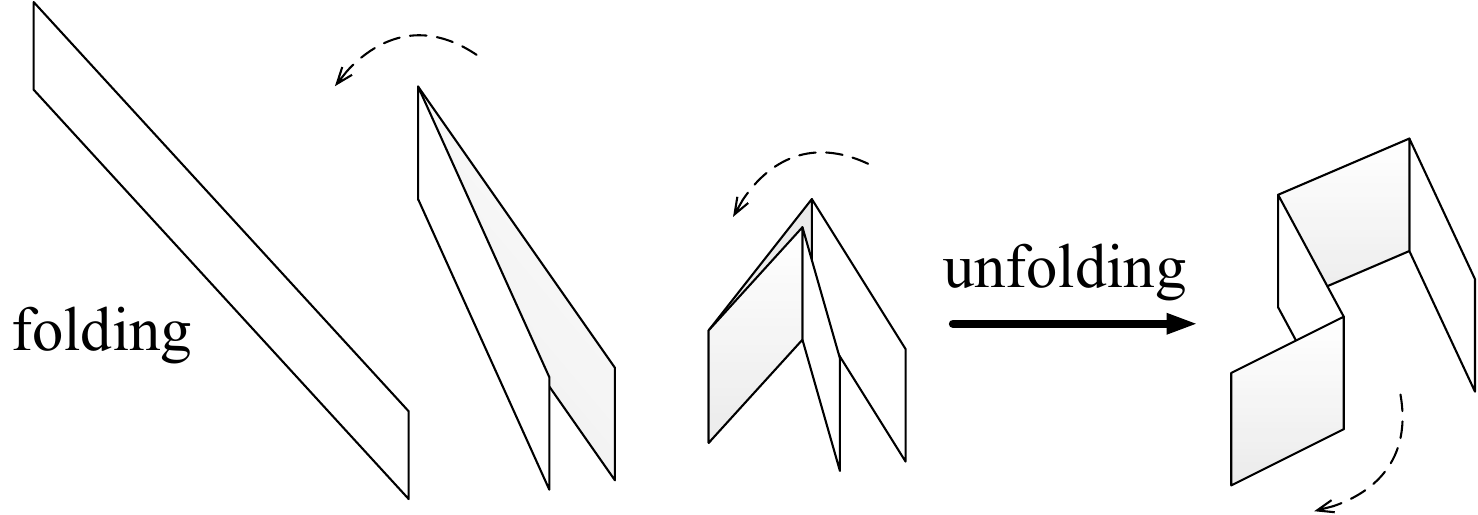}
\caption{}
\label{subfig:paper-folding}
\end{subfigure}
\begin{subfigure}[b]{0.3\textwidth}
\centering
\includegraphics[width=0.5\textwidth]{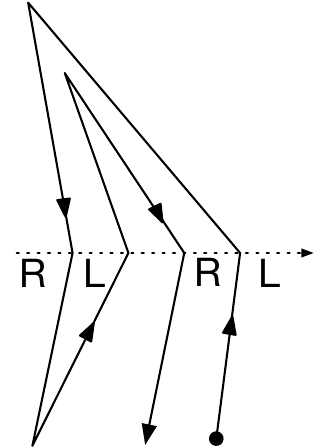}
\caption{\label{subfig:crease-making}}
\end{subfigure}
\caption{Dragon curve creation and paper folding.}
\label{fig:paper-folding}
\end{figure}

We may observe that the two approaches do generate the same curves, modulo resizing and rotation.
It becomes intuitive once we see that the dragon curve is the shape generated by folding paper repeatedly: take a strip of paper and keep folding it in half, as shown on the left of Figure~\ref{fig:paper-folding}(a).
The shape we get after we unfold the paper is the dragon curve ---
shown on the right of Figure~\ref{fig:paper-folding}(a) is the order 2 dragon curve, so the unfolding corresponds to copying and rotation in the \emph{mirroring} approach.
Shown in Figure~\ref{fig:paper-folding}(b) is what happens when we fold the order 2 paper once more.
As we can see, the new creases we encounter along the direction of the curve are
a left turn, a right turn, and a left turn... and so on --- we \emph{interleave} left and right turns into an existing curve, hence the name.
But how do we prove the equivalence?

Indeed, we would like to formally prove the equivalence of the two constructions.
This pearl is dedicated to such a proof.
The pearl is intended for readers who are interested in formal proofs and reasoning about
functional programs, who appreciate concepts such as proof by induction, folds,
and naturality.
We use a Haskell-like notation but the ideas carry over to other functional languages
(the use of infinite lists, for example, is merely for conceptual simplicity and not a necessity).
The pearl itself focuses on manual, equational reasoning, while Agda proofs that verify all the details are submitted to the journal as supplementary material.

\section{Formalisation}

To reason about the dragon curve, we first have to represent it formally.
A natural way to represent a curve is by a sequence of turns.
Let \ensuremath{\Conid{L}} denote a left (anticlockwise) turn and \ensuremath{\Conid{R}} a right (clockwise) turn:%
\footnote{Since it helps to know which lists are finite and which are not,
we denote the type of finite, inductive lists by \ensuremath{\Conid{List}} and infinite ones by \ensuremath{\Conid{Stream}}.
In Haskell both are represented by the same type, while in Agda they are not.}
\begin{hscode}\SaveRestoreHook
\column{B}{@{}>{\hspre}l<{\hspost}@{}}%
\column{13}{@{}>{\hspre}l<{\hspost}@{}}%
\column{26}{@{}>{\hspre}l<{\hspost}@{}}%
\column{E}{@{}>{\hspre}l<{\hspost}@{}}%
\>[B]{}\mathbf{type}\;\Conid{Curve}{}\<[13]%
\>[13]{}\mathrel{=}\Conid{List}\;\Conid{Turn}{}\<[26]%
\>[26]{}~~,{}\<[E]%
\\
\>[B]{}\mathbf{data}\;\Conid{Turn}{}\<[13]%
\>[13]{}\mathrel{=}\Conid{L}\mid \Conid{R}{}\<[26]%
\>[26]{}~~.{}\<[E]%
\ColumnHook
\end{hscode}\resethooks
The order \ensuremath{\mathrm{0}} dragon curve, a straight line with no turns, is represented by the empty list \ensuremath{[\mskip1.5mu \mskip1.5mu]}.
Curves of orders \ensuremath{\mathrm{1}} to \ensuremath{\mathrm{4}} are respectively represented by
\ensuremath{\Conid{L}}, \ensuremath{\Conid{LLR}}, \ensuremath{\Conid{LLRLLRR}}, and \ensuremath{\Conid{LLRLLRRLLLRRLRR}}.

Let \ensuremath{\Conid{Nat}} be the type of natural numbers.
The mirroring construction can be defined as:%
\footnote{In our definitions we use \ensuremath{\mathrm{1}\mathbin{+}\Varid{n}} as a pattern,
to make explicit the close connection between natural numbers and lists.}
\begin{hscode}\SaveRestoreHook
\column{B}{@{}>{\hspre}l<{\hspost}@{}}%
\column{4}{@{}>{\hspre}l<{\hspost}@{}}%
\column{16}{@{}>{\hspre}l<{\hspost}@{}}%
\column{E}{@{}>{\hspre}l<{\hspost}@{}}%
\>[B]{}\Varid{dragonM}\mathbin{::}\Conid{Nat}\to \Conid{Curve}{}\<[E]%
\\
\>[B]{}\Varid{dragonM}\;\mathrm{0}{}\<[16]%
\>[16]{}\mathrel{=}[\mskip1.5mu \mskip1.5mu]{}\<[E]%
\\
\>[B]{}\Varid{dragonM}\;(\mathrm{1}\mathbin{+}\Varid{n}){}\<[16]%
\>[16]{}\mathrel{=}\Varid{ts}\mathbin{{+}\mskip-8mu{+}}[\mskip1.5mu \Conid{L}\mskip1.5mu]\mathbin{{+}\mskip-8mu{+}}\Varid{map}\;\Varid{inv}\;(\Varid{reverse}\;\Varid{ts})~~,{}\<[E]%
\\
\>[B]{}\hsindent{4}{}\<[4]%
\>[4]{}\mathbf{where}\;\Varid{ts}\mathrel{=}\Varid{dragonM}\;\Varid{n}~~.{}\<[E]%
\ColumnHook
\end{hscode}\resethooks
The curve of order \ensuremath{\mathrm{1}\mathbin{+}\Varid{n}} starts with the same turns as the curve of order \ensuremath{\Varid{n}},
followed by an \ensuremath{\Conid{L}}, indicating the 90 degree clockwise rotation.
We then trace the same order \ensuremath{\Varid{n}} curve backwards (hence the \ensuremath{\Varid{reverse}}),
but because we are going backwards, each \ensuremath{\Conid{L}} becomes an \ensuremath{\Conid{R}}, and vice versa.
The function \ensuremath{\Varid{inv}} that inverts the turns is defined by:
\begin{hscode}\SaveRestoreHook
\column{B}{@{}>{\hspre}l<{\hspost}@{}}%
\column{8}{@{}>{\hspre}l<{\hspost}@{}}%
\column{E}{@{}>{\hspre}l<{\hspost}@{}}%
\>[B]{}\Varid{inv}\mathbin{::}\Conid{Turn}\to \Conid{Turn}{}\<[E]%
\\
\>[B]{}\Varid{inv}\;\Conid{L}{}\<[8]%
\>[8]{}\mathrel{=}\Conid{R}{}\<[E]%
\\
\>[B]{}\Varid{inv}\;\Conid{R}{}\<[8]%
\>[8]{}\mathrel{=}\Conid{L}~~.{}\<[E]%
\ColumnHook
\end{hscode}\resethooks

Now consider the interleaving approach.
With our representation, the act of `adding a crease to each line segment' is
performed by inserting alternating \ensuremath{\Conid{L}}'s and \ensuremath{\Conid{R}}'s into a sequence of turns.
We let \ensuremath{\Varid{alts}\;\Varid{t}} generate a stream of alternating \ensuremath{\Varid{t}}'s and \ensuremath{\Varid{inv}\;\Varid{t}}'s,
with which we define streams \ensuremath{\Varid{lr}\mathrel{=}\Conid{L}\mathbin{:}\Conid{R}\mathbin{:}\Conid{L}\mathbin{:}\Conid{R}\mathbin{...}} and \ensuremath{\Varid{rl}\mathrel{=}\Conid{R}\mathbin{:}\Conid{L}\mathbin{:}\Conid{R}\mathbin{:}\Conid{L}\mathbin{...}}:
\begin{hscode}\SaveRestoreHook
\column{B}{@{}>{\hspre}l<{\hspost}@{}}%
\column{9}{@{}>{\hspre}l<{\hspost}@{}}%
\column{29}{@{}>{\hspre}l<{\hspost}@{}}%
\column{E}{@{}>{\hspre}l<{\hspost}@{}}%
\>[B]{}\Varid{alts}{}\<[9]%
\>[9]{}\mathbin{::}\Conid{Turn}\to \Conid{Stream}\;\Conid{Turn}{}\<[E]%
\\
\>[B]{}\Varid{alts}\;\Varid{t}{}\<[9]%
\>[9]{}\mathrel{=}\Varid{t}\mathbin{:}\Varid{alts}\;(\Varid{inv}\;\Varid{t}){}\<[29]%
\>[29]{}~~,{}\<[E]%
\ColumnHook
\end{hscode}\resethooks
with \ensuremath{\Varid{lr}\mathrel{=}\Varid{alts}\;\Conid{L}} and \ensuremath{\Varid{rl}\mathrel{=}\Varid{alts}\;\Conid{R}}.

\begin{figure}
\centering
\includegraphics[width=0.5\textwidth]{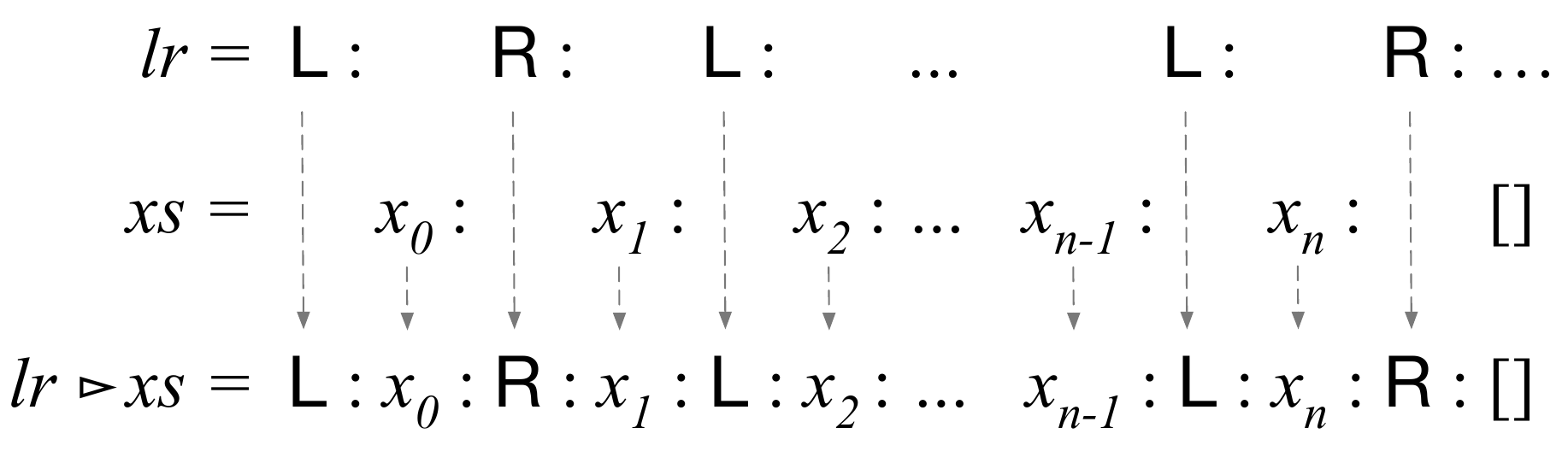}
\caption{Interleaving \ensuremath{\Varid{lr}\triangleright[\mskip1.5mu \Varid{x}_{0},\Varid{x}_{1}\mathinner{\ldotp\ldotp}x_n\mskip1.5mu]}.}
\label{fig:interleaving}
\end{figure}

Let the operator \ensuremath{(\triangleright)} denote `interleaving' of a stream and a list:
\begin{hscode}\SaveRestoreHook
\column{B}{@{}>{\hspre}l<{\hspost}@{}}%
\column{29}{@{}>{\hspre}l<{\hspost}@{}}%
\column{E}{@{}>{\hspre}l<{\hspost}@{}}%
\>[B]{}(\triangleright)\mathbin{::}\Conid{Stream}\;\Varid{a}\to \Conid{List}\;\Varid{a}\to \Conid{List}\;\Varid{a}{}\<[E]%
\\
\>[B]{}(\Varid{x}\mathbin{:}\Varid{xs})\triangleright[\mskip1.5mu \mskip1.5mu]{}\<[29]%
\>[29]{}\mathrel{=}[\mskip1.5mu \Varid{x}\mskip1.5mu]{}\<[E]%
\\
\>[B]{}(\Varid{x}\mathbin{:}\Varid{xs})\triangleright(\Varid{y}\mathbin{:}\Varid{ys}){}\<[29]%
\>[29]{}\mathrel{=}\Varid{x}\mathbin{:}\Varid{y}\mathbin{:}(\Varid{xs}\triangleright\Varid{ys})~~.{}\<[E]%
\ColumnHook
\end{hscode}\resethooks
It is a bit unusual that in \ensuremath{\Varid{xs}\triangleright\Varid{ys}}, it is assumed that \ensuremath{\Varid{xs}} is infinite, and when \ensuremath{\Varid{ys}} runs out, the sequence ends with the head of \ensuremath{\Varid{xs}}.
Figure~\ref{fig:interleaving} shows an example where we interleave the stream \ensuremath{\Varid{lr}} into \ensuremath{\Varid{xs}\mathrel{=}[\mskip1.5mu \Varid{x}_{0}\mathinner{\ldotp\ldotp}x_n\mskip1.5mu]}.
In the example the result ends with \ensuremath{\Conid{R}}, implying that \ensuremath{\Varid{xs}} has an odd number of elements --- we will see that the parity of the length of \ensuremath{\Varid{xs}} matters when we reason about \ensuremath{(\triangleright)}.

Given the ingredients, the interleaving approach is defined by:
\begin{hscode}\SaveRestoreHook
\column{B}{@{}>{\hspre}l<{\hspost}@{}}%
\column{16}{@{}>{\hspre}l<{\hspost}@{}}%
\column{E}{@{}>{\hspre}l<{\hspost}@{}}%
\>[B]{}\Varid{dragonI}\mathbin{::}\Conid{Nat}\to \Conid{Curve}{}\<[E]%
\\
\>[B]{}\Varid{dragonI}\;\mathrm{0}{}\<[16]%
\>[16]{}\mathrel{=}[\mskip1.5mu \mskip1.5mu]{}\<[E]%
\\
\>[B]{}\Varid{dragonI}\;(\mathrm{1}\mathbin{+}\Varid{n}){}\<[16]%
\>[16]{}\mathrel{=}\Varid{lr}\triangleright\Varid{dragonI}\;\Varid{n}~~.{}\<[E]%
\ColumnHook
\end{hscode}\resethooks
The task now is to prove that \ensuremath{\Varid{dragonM}\mathrel{=}\Varid{dragonI}}.

\section{The proof}

One might soon notice that both \ensuremath{\Varid{dragonM}} and \ensuremath{\Varid{dragonI}} are instances of folds on natural numbers, defined by:
\begin{hscode}\SaveRestoreHook
\column{B}{@{}>{\hspre}l<{\hspost}@{}}%
\column{11}{@{}>{\hspre}l<{\hspost}@{}}%
\column{18}{@{}>{\hspre}l<{\hspost}@{}}%
\column{E}{@{}>{\hspre}l<{\hspost}@{}}%
\>[B]{}\Varid{foldN}\mathbin{::}{}\<[11]%
\>[11]{}(\Varid{a}\to \Varid{a})\to \Varid{a}\to \Conid{Nat}\to \Varid{a}{}\<[E]%
\\
\>[B]{}\Varid{foldN}\;\Varid{f}\;\Varid{e}\;\mathrm{0}{}\<[18]%
\>[18]{}\mathrel{=}\Varid{e}{}\<[E]%
\\
\>[B]{}\Varid{foldN}\;\Varid{f}\;\Varid{e}\;(\mathrm{1}\mathbin{+}\Varid{n}){}\<[18]%
\>[18]{}\mathrel{=}\Varid{f}\;(\Varid{foldN}\;\Varid{f}\;\Varid{e}\;\Varid{n})~~,{}\<[E]%
\ColumnHook
\end{hscode}\resethooks
and equipped with the following \emph{universal property}:

\begin{hscode}\SaveRestoreHook
\column{B}{@{}>{\hspre}l<{\hspost}@{}}%
\column{3}{@{}>{\hspre}l<{\hspost}@{}}%
\column{18}{@{}>{\hspre}l<{\hspost}@{}}%
\column{E}{@{}>{\hspre}l<{\hspost}@{}}%
\>[3]{}\Varid{h}\mathrel{=}\Varid{foldN}\;\Varid{f}\;\Varid{e}{}\<[18]%
\>[18]{}~\Leftrightarrow~\Varid{h}\;\mathrm{0}\mathrel{=}\Varid{e}\,\mathrel{\wedge}\,(\forall\;\Varid{n}\mathbin{::}\Conid{Nat}\mathbin{:}\Varid{h}\;(\mathrm{1}\mathbin{+}\Varid{n})\mathrel{=}\Varid{f}\;(\Varid{h}\;\Varid{n}))~~.{}\<[E]%
\ColumnHook
\end{hscode}\resethooks
The two methods for constructing the dragon curve can be defined respectively by:
\begin{hscode}\SaveRestoreHook
\column{B}{@{}>{\hspre}l<{\hspost}@{}}%
\column{4}{@{}>{\hspre}l<{\hspost}@{}}%
\column{11}{@{}>{\hspre}l<{\hspost}@{}}%
\column{E}{@{}>{\hspre}l<{\hspost}@{}}%
\>[B]{}\Varid{dragonM}{}\<[11]%
\>[11]{}\mathrel{=}\Varid{foldN}\;\Varid{dm}\;[\mskip1.5mu \mskip1.5mu]~~,{}\<[E]%
\\
\>[B]{}\hsindent{4}{}\<[4]%
\>[4]{}\mathbf{where}\;\Varid{dm}\mathrel{=}\lambda \Varid{ts}\to \Varid{ts}\mathbin{{+}\mskip-8mu{+}}[\mskip1.5mu \Conid{L}\mskip1.5mu]\mathbin{{+}\mskip-8mu{+}}\Varid{map}\;\Varid{inv}\;(\Varid{reverse}\;\Varid{ts})~~,{}\<[E]%
\\
\>[B]{}\Varid{dragonI}{}\<[11]%
\>[11]{}\mathrel{=}\Varid{foldN}\;\Varid{di}\;[\mskip1.5mu \mskip1.5mu]~~,{}\<[E]%
\\
\>[B]{}\hsindent{4}{}\<[4]%
\>[4]{}\mathbf{where}\;\Varid{di}\mathrel{=}(\Varid{lr}\,\triangleright)~~.{}\<[E]%
\ColumnHook
\end{hscode}\resethooks
Thus, by the universal property, proving the equality \ensuremath{\Varid{dragonI}\mathrel{=}\Varid{dragonM}} is
equivalent to proving that \ensuremath{\Varid{dragonI}\;\mathrm{0}\mathrel{=}[\mskip1.5mu \mskip1.5mu]}, which holds by definition, and that
\begin{equation}
  \label{eq:dragonI-rec}
  \ensuremath{\Varid{dragonI}\;(\mathrm{1}\mathbin{+}\Varid{n})\mathrel{=}\Varid{dragonI}\;\Varid{n}\mathbin{{+}\mskip-8mu{+}}[\mskip1.5mu \Conid{L}\mskip1.5mu]\mathbin{{+}\mskip-8mu{+}}\Varid{map}\;\Varid{inv}\;(\Varid{reverse}\;(\Varid{dragonI}\;\Varid{n}))} \mbox{~~,}
\end{equation}
by choosing \ensuremath{\Varid{h}\mathrel{=}\Varid{dragonI}} and \ensuremath{\Varid{f}\mathrel{=}\Varid{dm}}.

We prove \eqref{eq:dragonI-rec} by induction on \ensuremath{\Varid{n}}.
The base case for \ensuremath{\mathrm{0}} is immediate.
The inductive case for \ensuremath{\mathrm{1}\mathbin{+}\Varid{n}} starts with:
\begin{hscode}\SaveRestoreHook
\column{B}{@{}>{\hspre}l<{\hspost}@{}}%
\column{7}{@{}>{\hspre}l<{\hspost}@{}}%
\column{E}{@{}>{\hspre}l<{\hspost}@{}}%
\>[7]{}\Varid{dragonI}\;(\mathrm{2}\mathbin{+}\Varid{n}){}\<[E]%
\\
\>[B]{}\mathrel{=}{}\<[7]%
\>[7]{}\mbox{\commentbegin  definition of \ensuremath{\Varid{dragonI}}  \commentend}{}\<[E]%
\\
\>[7]{}\Varid{lr}\triangleright\Varid{dragonI}\;(\mathrm{1}\mathbin{+}\Varid{n}){}\<[E]%
\\
\>[B]{}\mathrel{=}{}\<[7]%
\>[7]{}\mbox{\commentbegin  induction hypothesis  \commentend}{}\<[E]%
\\
\>[7]{}\Varid{lr}\triangleright(\Varid{dragonI}\;\Varid{n}\mathbin{{+}\mskip-8mu{+}}[\mskip1.5mu \Conid{L}\mskip1.5mu]\mathbin{{+}\mskip-8mu{+}}\Varid{map}\;\Varid{inv}\;(\Varid{reverse}\;(\Varid{dragonI}\;\Varid{n}))){}\<[E]%
\\
\>[B]{}\mathrel{=}{}\<[7]%
\>[7]{}\mathbin{...}{}\<[E]%
\ColumnHook
\end{hscode}\resethooks
The aim now is to bring \ensuremath{(\Varid{lr}\,\triangleright)} next to the two occurrences of \ensuremath{\Varid{dragonI}\;\Varid{n}} ---
once we create an \ensuremath{\Varid{lr}\triangleright\Varid{dragonI}\;\Varid{n}}, we can fold it back to \ensuremath{\Varid{dragonI}\;(\mathrm{1}\mathbin{+}\Varid{n})}.
For that we want to look at how \ensuremath{(\Varid{lr}\,\triangleright)} distributes over \ensuremath{(\mathbin{{+}\mskip-8mu{+}})}, \ensuremath{\Varid{map}}, \ensuremath{\Varid{reverse}}, etc.

Note that we have been thinking syntactically so far --- by letting the symbols do the work, we discovered that we need some form of distributivity.
We will now temporarily switch to semantics mode, thinking about what distributivity property we might actually have.
These properties will then be proved in a syntax-driven manner.

\subsubsection*{Distributivity}

Firstly, what happens when \ensuremath{(\Varid{lr}\,\triangleright)} encounters \ensuremath{(\mathbin{{+}\mskip-8mu{+}})}?
It turns out that for all \ensuremath{\Varid{xs}}, \ensuremath{\Varid{ys}}, and \ensuremath{\Varid{z}}, where \ensuremath{\Varid{xs}} has an odd length, \ensuremath{(\Varid{alts}\;\Varid{t}\,\triangleright)} distributes over \ensuremath{(\mathbin{{+}\mskip-8mu{+}})} in the way below:
\begin{equation}
\label{eq:interleave-cat}
\setlength{\jot}{0pt}
\begin{split}
    &  \ensuremath{\Varid{odd}\;(\Varid{length}\;\Varid{xs})\Rightarrow}\\
    &   \quad \ensuremath{\Varid{alts}\;\Varid{t}\triangleright(\Varid{xs}\mathbin{{+}\mskip-8mu{+}}[\mskip1.5mu \Varid{z}\mskip1.5mu]\mathbin{{+}\mskip-8mu{+}}\Varid{ys})\mathrel{=}(\Varid{alts}\;\Varid{t}\triangleright\Varid{xs})\mathbin{{+}\mskip-8mu{+}}[\mskip1.5mu \Varid{z}\mskip1.5mu]\mathbin{{+}\mskip-8mu{+}}(\Varid{alts}\;\Varid{t}\triangleright\Varid{ys})} \mbox{~~.}
\end{split}
\end{equation}
For some intuition towards \eqref{eq:interleave-cat}, try \ensuremath{\Varid{t}\mathbin{:=}\Conid{L}}, \ensuremath{\Varid{xs}\mathbin{:=}[\mskip1.5mu \Varid{a},\Varid{b},\Varid{c}\mskip1.5mu]}, and \ensuremath{\Varid{ys}\mathbin{:=}[\mskip1.5mu \Varid{d},\Varid{e}\mskip1.5mu]}:
\begin{hscode}\SaveRestoreHook
\column{B}{@{}>{\hspre}c<{\hspost}@{}}%
\column{BE}{@{}l@{}}%
\column{4}{@{}>{\hspre}l<{\hspost}@{}}%
\column{E}{@{}>{\hspre}l<{\hspost}@{}}%
\>[4]{}\Varid{lr}\triangleright([\mskip1.5mu \Varid{a},\Varid{b},\Varid{c}\mskip1.5mu]\mathbin{{+}\mskip-8mu{+}}[\mskip1.5mu \Varid{z}\mskip1.5mu]\mathbin{{+}\mskip-8mu{+}}[\mskip1.5mu \Varid{d},\Varid{e}\mskip1.5mu]){}\<[E]%
\\
\>[B]{}\mathrel{=}{}\<[BE]%
\>[4]{}[\mskip1.5mu \Conid{L},\Varid{a},\Conid{R},\Varid{b},\Conid{L},\Varid{c},\Conid{R},\Varid{z},\Conid{L},\Varid{d},\Conid{R},\Varid{e},\Conid{L}\mskip1.5mu]{}\<[E]%
\\
\>[B]{}\mathrel{=}{}\<[BE]%
\>[4]{}[\mskip1.5mu \Conid{L},\Varid{a},\Conid{R},\Varid{b},\Conid{L},\Varid{c},\Conid{R}\mskip1.5mu]\mathbin{{+}\mskip-8mu{+}}[\mskip1.5mu \Varid{z}\mskip1.5mu]\mathbin{{+}\mskip-8mu{+}}[\mskip1.5mu \Conid{L},\Varid{d},\Conid{R},\Varid{e},\Conid{L}\mskip1.5mu]{}\<[E]%
\\
\>[B]{}\mathrel{=}{}\<[BE]%
\>[4]{}(\Varid{lr}\triangleright[\mskip1.5mu \Varid{a},\Varid{b},\Varid{c}\mskip1.5mu])\mathbin{{+}\mskip-8mu{+}}[\mskip1.5mu \Varid{z}\mskip1.5mu]\mathbin{{+}\mskip-8mu{+}}(\Varid{lr}\triangleright[\mskip1.5mu \Varid{d},\Varid{e}\mskip1.5mu])~~.{}\<[E]%
\ColumnHook
\end{hscode}\resethooks
The proof of \eqref{eq:interleave-cat} is an induction on odd-length%
\footnote{In the Agda proof we define a predicate for odd-length lists and perform induction on that predicate.
For completeness, details are given in Appendix~\ref{sec:proof-distributivity}.}
\ensuremath{\Varid{xs}}: the base case is a singleton list \ensuremath{[\mskip1.5mu \Varid{x}\mskip1.5mu]}, and the inductive case is \ensuremath{\Varid{x}\mathbin{:}\Varid{y}\mathbin{:}\Varid{xs}}.

Secondly, by naturality of \ensuremath{(\triangleright)} we have that for all \ensuremath{\Varid{xs}}, \ensuremath{\Varid{zs}}, and \ensuremath{\Varid{f}},
\begin{equation*}
   \ensuremath{\Varid{map}\;\Varid{f}\;\Varid{zs}\triangleright\Varid{map}\;\Varid{f}\;\Varid{xs}\mathrel{=}\Varid{map}\;\Varid{f}\;(\Varid{zs}\triangleright\Varid{xs})~~.}
\end{equation*}
Together with \ensuremath{\Varid{inv}\circo\Varid{inv}\mathrel{=}\Varid{id}}, we have the following property showing what happens
when \ensuremath{(\Varid{zs}\,\triangleright)} encounters \ensuremath{\Varid{map}\;\Varid{inv}}:
\begin{equation}
\label{eq:interleave-inv}
 \ensuremath{\Varid{zs}\triangleright\Varid{map}\;\Varid{inv}\;\Varid{xs}\mathrel{=}\Varid{map}\;\Varid{inv}\;(\Varid{map}\;\Varid{inv}\;\Varid{zs}\triangleright\Varid{xs})}
 \mbox{~~.}
\end{equation}
Finally, for all \ensuremath{\Varid{xs}} having an odd length, this is how \ensuremath{\Varid{reverse}} distributes over \ensuremath{(\triangleright)}:
\begin{equation}
\label{eq:interleave-reverse}
\setlength{\jot}{0pt}
\begin{split}
 & \ensuremath{\Varid{odd}\;(\Varid{length}\;\Varid{xs})\Rightarrow}\\
 & \quad \ensuremath{\Varid{reverse}\;(\Varid{alts}\;\Varid{t}\triangleright\Varid{xs})\mathrel{=}\Varid{alts}\;(\Varid{inv}\;\Varid{t})\triangleright\Varid{reverse}\;\Varid{xs}} \mbox{~~.}
\end{split}
\end{equation}
Again, to understand \eqref{eq:interleave-reverse},
try an example \ensuremath{\Varid{t}\mathbin{:=}\Conid{L}} and \ensuremath{\Varid{xs}\mathbin{:=}[\mskip1.5mu \Varid{a},\Varid{b},\Varid{c}\mskip1.5mu]}:
\begin{hscode}\SaveRestoreHook
\column{B}{@{}>{\hspre}c<{\hspost}@{}}%
\column{BE}{@{}l@{}}%
\column{4}{@{}>{\hspre}l<{\hspost}@{}}%
\column{E}{@{}>{\hspre}l<{\hspost}@{}}%
\>[4]{}\Varid{reverse}\;(\Varid{lr}\triangleright[\mskip1.5mu \Varid{a},\Varid{b},\Varid{c}\mskip1.5mu]){}\<[E]%
\\
\>[B]{}\mathrel{=}{}\<[BE]%
\>[4]{}\Varid{reverse}\;[\mskip1.5mu \Conid{L},\Varid{a},\Conid{R},\Varid{b},\Conid{L},\Varid{c},\Conid{R}\mskip1.5mu]{}\<[E]%
\\
\>[B]{}\mathrel{=}{}\<[BE]%
\>[4]{}[\mskip1.5mu \Conid{R},\Varid{c},\Conid{L},\Varid{b},\Conid{R},\Varid{a},\Conid{L}\mskip1.5mu]{}\<[E]%
\\
\>[B]{}\mathrel{=}{}\<[BE]%
\>[4]{}\Varid{rl}\triangleright[\mskip1.5mu \Varid{c},\Varid{b},\Varid{a}\mskip1.5mu]~~.{}\<[E]%
\ColumnHook
\end{hscode}\resethooks
Proof of \eqref{eq:interleave-reverse} is again an induction on odd-length \ensuremath{\Varid{xs}}.
We will need \eqref{eq:interleave-cat}, as well as properties including associativity of \ensuremath{(\mathbin{{+}\mskip-8mu{+}})} and that \ensuremath{\Varid{reverse}} preserves lengths.
Details are given in Appendix~\ref{sec:proof-distributivity}.

\subsubsection*{Back to the proof of \eqref{eq:dragonI-rec}}

As mentioned before, we prove \eqref{eq:dragonI-rec} by induction,
and the case for \ensuremath{\mathrm{0}} is immediate.
For the inductive case,
the calculation below has a clear goal --- to move \ensuremath{(\Varid{lr}\,\triangleright)} next to
\ensuremath{\Varid{dragonI}\;\Varid{n}} using the properties we mentioned just now, to fold them back to \ensuremath{\Varid{dragonI}\;(\mathrm{1}\mathbin{+}\Varid{n})}.
To use \eqref{eq:interleave-cat} and \eqref{eq:interleave-reverse}, we need \ensuremath{\Varid{length}\;(\Varid{dragonI}\;\Varid{n})} to be an odd number.
This is true for \ensuremath{\Varid{n}\mathbin{>}\mathrm{0}}, since \ensuremath{(\triangleright)} always generates odd-length lists, therefore the calculation below is for \ensuremath{\Varid{n}\mathbin{>}\mathrm{0}}:
\begin{hscode}\SaveRestoreHook
\column{B}{@{}>{\hspre}l<{\hspost}@{}}%
\column{5}{@{}>{\hspre}l<{\hspost}@{}}%
\column{8}{@{}>{\hspre}l<{\hspost}@{}}%
\column{E}{@{}>{\hspre}l<{\hspost}@{}}%
\>[5]{}\Varid{dragonI}\;(\mathrm{2}\mathbin{+}\Varid{n}){}\<[E]%
\\
\>[B]{}\mathrel{=}{}\<[8]%
\>[8]{}\mbox{\commentbegin  definition of \ensuremath{\Varid{dragonI}}  \commentend}{}\<[E]%
\\
\>[B]{}\hsindent{5}{}\<[5]%
\>[5]{}\Varid{lr}\triangleright\Varid{dragonI}\;(\mathrm{1}\mathbin{+}\Varid{n}){}\<[E]%
\\
\>[B]{}\mathrel{=}{}\<[8]%
\>[8]{}\mbox{\commentbegin  induction hypothesis  \commentend}{}\<[E]%
\\
\>[B]{}\hsindent{5}{}\<[5]%
\>[5]{}\Varid{lr}\triangleright(\Varid{dragonI}\;\Varid{n}\mathbin{{+}\mskip-8mu{+}}[\mskip1.5mu \Conid{L}\mskip1.5mu]\mathbin{{+}\mskip-8mu{+}}\Varid{map}\;\Varid{inv}\;(\Varid{reverse}\;(\Varid{dragonI}\;\Varid{n}))){}\<[E]%
\\
\>[B]{}\mathrel{=}{}\<[8]%
\>[8]{}\mbox{\commentbegin  \ensuremath{\Varid{odd}\;(\Varid{length}\;(\Varid{dragonI}\;\Varid{n}))} if \ensuremath{\Varid{n}\mathbin{>}\mathrm{0}}, by \eqref{eq:interleave-cat}  \commentend}{}\<[E]%
\\
\>[B]{}\hsindent{5}{}\<[5]%
\>[5]{}(\Varid{lr}\triangleright\Varid{dragonI}\;\Varid{n})\mathbin{{+}\mskip-8mu{+}}[\mskip1.5mu \Conid{L}\mskip1.5mu]\mathbin{{+}\mskip-8mu{+}}(\Varid{lr}\triangleright\Varid{map}\;\Varid{inv}\;(\Varid{reverse}\;(\Varid{dragonI}\;\Varid{n}))){}\<[E]%
\\
\>[B]{}\mathrel{=}{}\<[8]%
\>[8]{}\mbox{\commentbegin  by \eqref{eq:interleave-inv}  \commentend}{}\<[E]%
\\
\>[B]{}\hsindent{5}{}\<[5]%
\>[5]{}(\Varid{lr}\triangleright\Varid{dragonI}\;\Varid{n})\mathbin{{+}\mskip-8mu{+}}[\mskip1.5mu \Conid{L}\mskip1.5mu]\mathbin{{+}\mskip-8mu{+}}\Varid{map}\;\Varid{inv}\;(\Varid{map}\;\Varid{inv}\;\Varid{lr}\triangleright\Varid{reverse}\;(\Varid{dragonI}\;\Varid{n})){}\<[E]%
\\
\>[B]{}\mathrel{=}{}\<[8]%
\>[8]{}\mbox{\commentbegin  since \ensuremath{\Varid{map}\;\Varid{inv}\;\Varid{lr}\mathrel{=}\Varid{rl}}  \commentend}{}\<[E]%
\\
\>[B]{}\hsindent{5}{}\<[5]%
\>[5]{}(\Varid{lr}\triangleright\Varid{dragonI}\;\Varid{n})\mathbin{{+}\mskip-8mu{+}}[\mskip1.5mu \Conid{L}\mskip1.5mu]\mathbin{{+}\mskip-8mu{+}}\Varid{map}\;\Varid{inv}\;(\Varid{rl}\triangleright\Varid{reverse}\;(\Varid{dragonI}\;\Varid{n})){}\<[E]%
\\
\>[B]{}\mathrel{=}{}\<[8]%
\>[8]{}\mbox{\commentbegin  \ensuremath{\Varid{odd}\;(\Varid{length}\;(\Varid{dragonI}\;\Varid{n}))} if \ensuremath{\Varid{n}\mathbin{>}\mathrm{0}}, by \eqref{eq:interleave-reverse}  \commentend}{}\<[E]%
\\
\>[B]{}\hsindent{5}{}\<[5]%
\>[5]{}(\Varid{lr}\triangleright\Varid{dragonI}\;\Varid{n})\mathbin{{+}\mskip-8mu{+}}[\mskip1.5mu \Conid{L}\mskip1.5mu]\mathbin{{+}\mskip-8mu{+}}\Varid{map}\;\Varid{inv}\;(\Varid{reverse}\;(\Varid{lr}\triangleright\Varid{dragonI}\;\Varid{n})){}\<[E]%
\\
\>[B]{}\mathrel{=}{}\<[8]%
\>[8]{}\mbox{\commentbegin  definition of \ensuremath{\Varid{dragonI}}  \commentend}{}\<[E]%
\\
\>[B]{}\hsindent{5}{}\<[5]%
\>[5]{}\Varid{dragonI}\;(\mathrm{1}\mathbin{+}\Varid{n})\mathbin{{+}\mskip-8mu{+}}[\mskip1.5mu \Conid{L}\mskip1.5mu]\mathbin{{+}\mskip-8mu{+}}\Varid{map}\;\Varid{inv}\;(\Varid{reverse}\;(\Varid{dragonI}\;(\mathrm{1}\mathbin{+}\Varid{n})))~~.{}\<[E]%
\ColumnHook
\end{hscode}\resethooks
The case for \ensuremath{\Varid{n}\mathrel{=}\mathrm{0}} (that is, \ensuremath{\Varid{dragonI}\;\mathrm{2}\mathrel{=}\Varid{dragonI}\;\mathrm{1}\mathbin{{+}\mskip-8mu{+}}[\mskip1.5mu \Conid{L}\mskip1.5mu]\mathbin{{+}\mskip-8mu{+}}\Varid{map}\;\Varid{inv}\;(\Varid{reverse}\;(\Varid{dragonI}\;\mathrm{1}))}) holds simply by definitions.
We have therefore proved \eqref{eq:dragonI-rec} for all \ensuremath{\Varid{n}}.

This is an illustrative example of a symbol-driven proof development.
Distributivity is your friend: the calculation was motivated by wanting to distribute \ensuremath{(\Varid{lr}\,\triangleright)} inwards.
That motivated us to temporarily think semantically and led us to properties \eqref{eq:interleave-cat}-\eqref{eq:interleave-reverse}.
During the development it may turn out that certain proofs only work given certain preconditions (e.g. \ensuremath{\Varid{length}\;(\Varid{dragonI}\;\Varid{n})} being odd, implied by \ensuremath{\Varid{n}\mathbin{>}\mathrm{0}}),
which prompts us to separately consider cases when the preconditions do not hold.

Note that the core of the proof (between the third step and the penultimate step), and the additional \ensuremath{\Varid{n}\mathrel{=}\mathrm{0}} case, can be abstracted as:
\begin{equation}
 \begin{split}
 & \ensuremath{\Varid{xs}\mathrel{=}[\mskip1.5mu \mskip1.5mu]\mathrel{\vee}\Varid{odd}\;(\Varid{length}\;\Varid{xs})\Rightarrow}\\
  &\qquad \ensuremath{\Varid{alts}\;\Varid{t}\triangleright(\Varid{xs}\mathbin{{+}\mskip-8mu{+}}[\mskip1.5mu \Varid{m}\mskip1.5mu]\mathbin{{+}\mskip-8mu{+}}\Varid{map}\;\Varid{inv}\;(\Varid{reverse}\;\Varid{xs}))\mathrel{=}}\\
  &\qquad\qquad  \ensuremath{(\Varid{alts}\;\Varid{t}\triangleright\Varid{xs})\mathbin{{+}\mskip-8mu{+}}[\mskip1.5mu \Varid{m}\mskip1.5mu]\mathbin{{+}\mskip-8mu{+}}\Varid{map}\;\Varid{inv}\;(\Varid{reverse}\;(\Varid{alts}\;\Varid{t}\triangleright\Varid{xs}))} \mbox{~~,}
\end{split}
\label{eq:lr-interleave-distr}
\end{equation}
for all \ensuremath{\Varid{t}} and \ensuremath{\Varid{m}}. We will see this equality again later.

\section{Generalisation}

It is a natural question to ask: what happens if we are allowed, in the mirroring construction, to rotate anticlockwise as well?
Indeed, such a generalisation has been studied by, for example, \citet{DavisKnuth:70:Number} and \citet{Tabachnikov:14:Dragon}.

\begin{figure}
\begin{subfigure}[b]{0.4\textwidth}
\centering
\def\svgwidth{4cm}
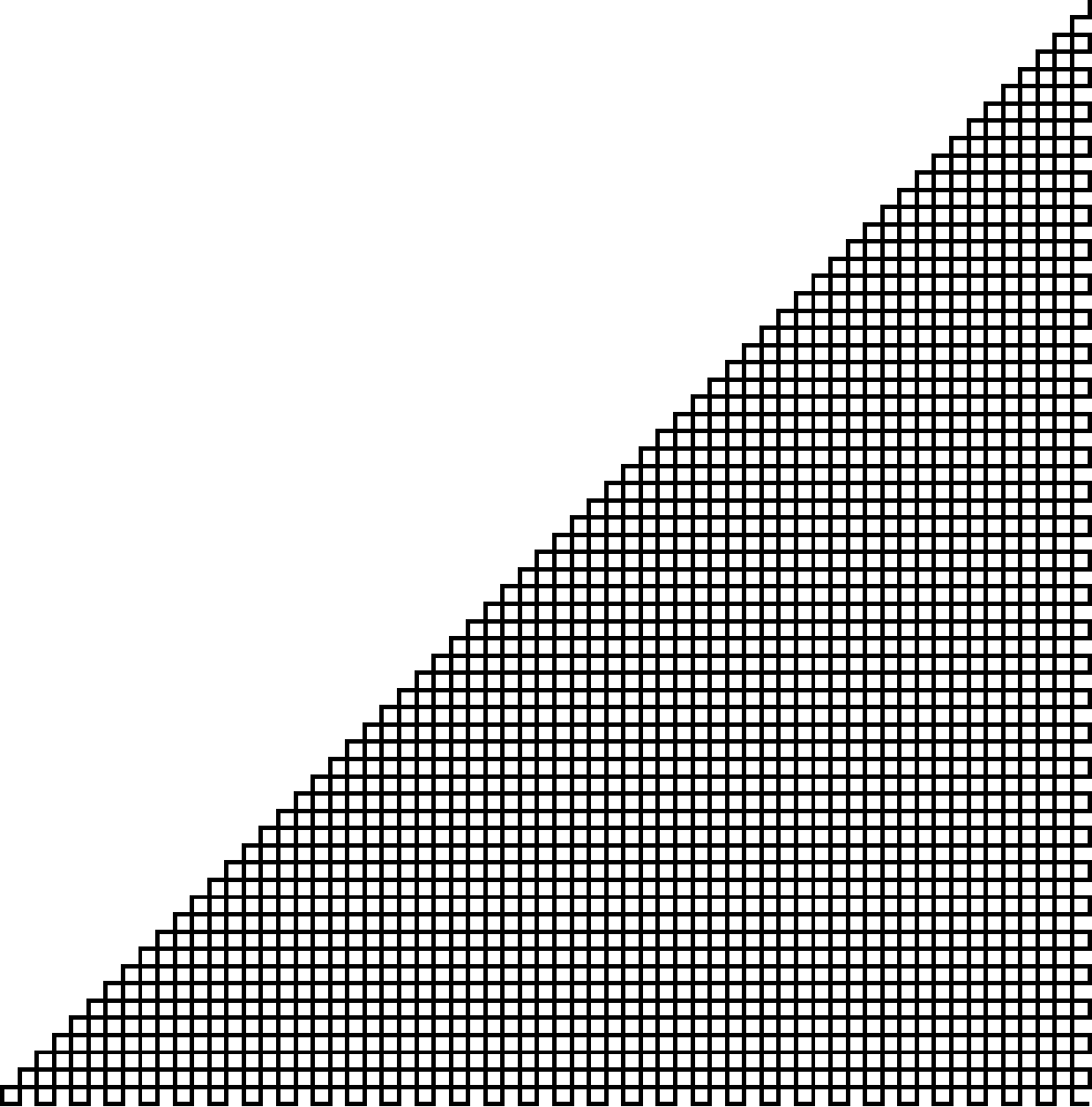
\caption{\ensuremath{[\mskip1.5mu \Conid{C},\Conid{A},\Conid{C},\Conid{A},\Conid{C},\Conid{A},\Conid{C},\Conid{A},\Conid{C},\Conid{A},\Conid{C},\Conid{A}\mskip1.5mu]}}
\label{subfig:triangle}
\end{subfigure}
\begin{subfigure}[b]{0.5\textwidth}
\centering
\def\svgwidth{5cm}
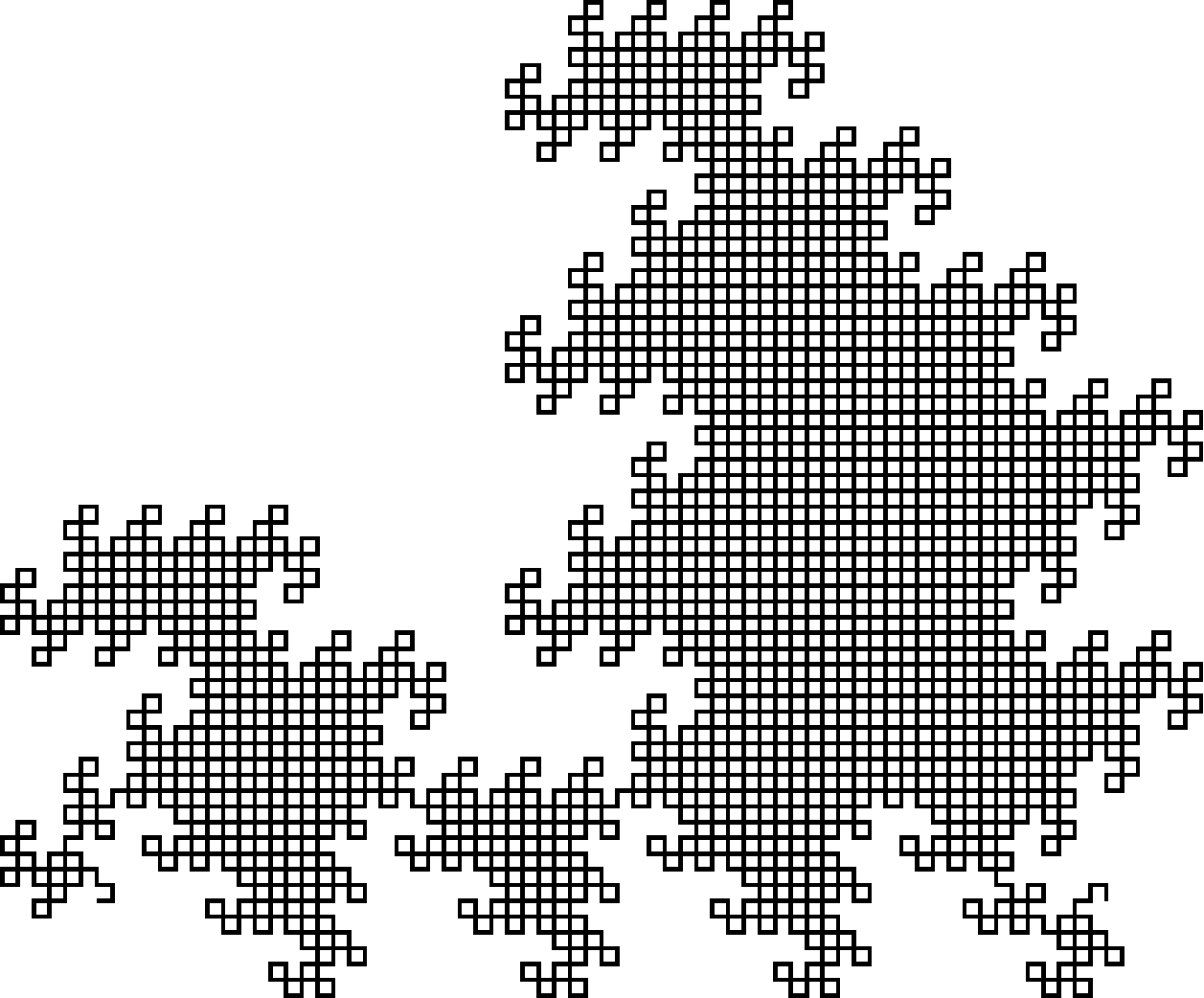
\caption{\ensuremath{[\mskip1.5mu \Conid{A},\Conid{C},\Conid{A},\Conid{A},\Conid{C},\Conid{A},\Conid{A},\Conid{A},\Conid{C},\Conid{C},\Conid{C},\Conid{C}\mskip1.5mu]}}
\label{subfig:mama}
\end{subfigure}
\caption{Generalised dragon curves.}
\label{fig:dragon-generalised}
\end{figure}

The generalised construction takes a list of instructions on which side to rotate, denoted by the type \ensuremath{\Conid{Rot}}:%
\footnote{Some authors reuse \ensuremath{\Conid{Turn}} here, but we think it is clearer to use a different type.}
\begin{hscode}\SaveRestoreHook
\column{B}{@{}>{\hspre}l<{\hspost}@{}}%
\column{E}{@{}>{\hspre}l<{\hspost}@{}}%
\>[B]{}\mathbf{data}\;\Conid{Rot}\mathrel{=}\Conid{C}\mid \Conid{A}~~,{}\<[E]%
\ColumnHook
\end{hscode}\resethooks
where \ensuremath{\Conid{C}} denotes clockwise and \ensuremath{\Conid{A}} anticlockwise.
The function \ensuremath{\Varid{dragonM}}, which was a fold on \ensuremath{\Conid{Nat}}, is generalised to a \ensuremath{\Varid{foldr}} taking \ensuremath{\Conid{List}\;\Conid{Rot}} as its input:
\begin{hscode}\SaveRestoreHook
\column{B}{@{}>{\hspre}l<{\hspost}@{}}%
\column{3}{@{}>{\hspre}l<{\hspost}@{}}%
\column{10}{@{}>{\hspre}l<{\hspost}@{}}%
\column{22}{@{}>{\hspre}l<{\hspost}@{}}%
\column{E}{@{}>{\hspre}l<{\hspost}@{}}%
\>[B]{}\Varid{dragonM}\mathbin{::}\Conid{List}\;\Conid{Rot}\to \Conid{Curve}{}\<[E]%
\\
\>[B]{}\Varid{dragonM}\mathrel{=}\Varid{foldr}\;(\oplus_{M})\;[\mskip1.5mu \mskip1.5mu]~~,{}\<[E]%
\\
\>[B]{}\hsindent{3}{}\<[3]%
\>[3]{}\mathbf{where}\;{}\<[10]%
\>[10]{}\Conid{C}\oplus_{M}\Varid{ts}{}\<[22]%
\>[22]{}\mathrel{=}\Varid{ts}\mathbin{{+}\mskip-8mu{+}}[\mskip1.5mu \Conid{L}\mskip1.5mu]\mathbin{{+}\mskip-8mu{+}}\Varid{map}\;\Varid{inv}\;(\Varid{reverse}\;\Varid{ts}){}\<[E]%
\\
\>[10]{}\Conid{A}\oplus_{M}\Varid{ts}{}\<[22]%
\>[22]{}\mathrel{=}\Varid{ts}\mathbin{{+}\mskip-8mu{+}}[\mskip1.5mu \Conid{R}\mskip1.5mu]\mathbin{{+}\mskip-8mu{+}}\Varid{map}\;\Varid{inv}\;(\Varid{reverse}\;\Varid{ts})~~.{}\<[E]%
\ColumnHook
\end{hscode}\resethooks
When the instruction is \ensuremath{\Conid{C}}, we insert \ensuremath{\Conid{L}} in the middle;
when the instruction is \ensuremath{\Conid{A}}, we insert \ensuremath{\Conid{R}} instead.
It is known that by alternating \ensuremath{\Conid{C}} and \ensuremath{\Conid{A}} we get a triangle, as seen in Figure~\ref{fig:dragon-generalised}(a);
in Figure~\ref{fig:dragon-generalised}(b) is a curve coined by \citet{DavisKnuth:70:Number} as `mama, baby, and papa', generated by \ensuremath{[\mskip1.5mu \Conid{A},\Conid{C},\Conid{A},\Conid{A},\Conid{C},\Conid{A},\Conid{A},\Conid{A},\Conid{C},\Conid{C},\Conid{C},\Conid{C}\mskip1.5mu]}.

The function \ensuremath{\Varid{dragonI}} can be generalised to take instructions as well.
However, with paper folding and unfolding being dual operations, \ensuremath{\Varid{dragonI}} applies the instructions in an order opposite to that of \ensuremath{\Varid{dragonM}}.
Therefore, \ensuremath{\Varid{dragonI}} is generalised to a \ensuremath{\Varid{foldl}}:
\begin{hscode}\SaveRestoreHook
\column{B}{@{}>{\hspre}l<{\hspost}@{}}%
\column{3}{@{}>{\hspre}l<{\hspost}@{}}%
\column{10}{@{}>{\hspre}l<{\hspost}@{}}%
\column{22}{@{}>{\hspre}l<{\hspost}@{}}%
\column{28}{@{}>{\hspre}l<{\hspost}@{}}%
\column{E}{@{}>{\hspre}l<{\hspost}@{}}%
\>[B]{}\Varid{dragonI}\mathbin{::}\Conid{List}\;\Conid{Rot}\to \Conid{Curve}{}\<[E]%
\\
\>[B]{}\Varid{dragonI}\mathrel{=}\Varid{foldl}\;(\oplus_{I})\;[\mskip1.5mu \mskip1.5mu]~~,{}\<[E]%
\\
\>[B]{}\hsindent{3}{}\<[3]%
\>[3]{}\mathbf{where}\;{}\<[10]%
\>[10]{}\Varid{ts}\oplus_{I}\Conid{C}{}\<[22]%
\>[22]{}\mathrel{=}\Varid{lr}{}\<[28]%
\>[28]{}\triangleright\Varid{ts}{}\<[E]%
\\
\>[10]{}\Varid{ts}\oplus_{I}\Conid{A}{}\<[22]%
\>[22]{}\mathrel{=}\Varid{rl}{}\<[28]%
\>[28]{}\triangleright\Varid{ts}~~.{}\<[E]%
\ColumnHook
\end{hscode}\resethooks
When we encounter a \ensuremath{\Conid{C}}, \ensuremath{\Varid{lr}} is interleaved into \ensuremath{\Varid{ts}}; when the instruction is \ensuremath{\Conid{A}}, it is \ensuremath{\Varid{rl}} that is interleaved.
The task now is, again, to prove that \ensuremath{\Varid{dragonM}\mathrel{=}\Varid{dragonI}}.

To prove that a \ensuremath{\Varid{foldr}} equals a \ensuremath{\Varid{foldl}}, recall the \emph{second duality theorem} (\citealt[p.~128]{Bird:98:Introduction}; see also \citealt[p.~124]{Bird:14:Thinking}):
\begin{sec-duality}{\rm \ensuremath{\Varid{foldr}\;(\oplus)\;\Varid{e}\;\Varid{xs}\mathrel{=}\Varid{foldl}\;(\otimes)\;\Varid{e}\;\Varid{xs}} for all finite lists \ensuremath{\Varid{xs}} if}
\begin{align}
\ensuremath{\Varid{x}\oplus\Varid{e}} &= \ensuremath{\Varid{e}\otimes\Varid{x}}  \label{eq:2nd-dual-base}\\
\ensuremath{\Varid{x}\oplus(\Varid{y}\otimes\Varid{z})} &= \ensuremath{(\Varid{x}\oplus\Varid{y})\otimes\Varid{z}} \mbox{~~.} \label{eq:2nd-dual-assoc}
\end{align}
\end{sec-duality}%
\noindent We will need a finer version of the theorem --- that \eqref{eq:2nd-dual-assoc} needs to hold only for those \ensuremath{\Varid{y}} in the range of \ensuremath{\Varid{foldl}\;(\otimes)\;\Varid{e}}.
Note that, given the constraint, the theorem becomes an equivalence ---
that is, \ensuremath{\Varid{foldr}\;(\oplus)\;\Varid{e}\;\Varid{xs}\mathrel{=}\Varid{foldl}\;(\otimes)\;\Varid{e}\;\Varid{xs}} \emph{if and only if} \eqref{eq:2nd-dual-base} and \eqref{eq:2nd-dual-assoc} hold for \ensuremath{\Varid{y}} in the range of \ensuremath{\Varid{foldl}\;(\otimes)\;\Varid{e}}.

To prove that \ensuremath{\Varid{dragonM}\mathrel{=}\Varid{dragonI}} using the theorem, our instance of \eqref{eq:2nd-dual-base}, \ensuremath{\Varid{r}\oplus_{M}[\mskip1.5mu \mskip1.5mu]\mathrel{=}[\mskip1.5mu \mskip1.5mu]\oplus_{I}\Varid{r}}
holds immediately: both sides simplify to \ensuremath{[\mskip1.5mu \Conid{L}\mskip1.5mu]} if \ensuremath{\Varid{r}\mathrel{=}\Conid{C}} or to \ensuremath{[\mskip1.5mu \Conid{R}\mskip1.5mu]} if \ensuremath{\Varid{r}\mathrel{=}\Conid{A}}.
Requirement \eqref{eq:2nd-dual-assoc} instantiates to:
\begin{hscode}\SaveRestoreHook
\column{B}{@{}>{\hspre}l<{\hspost}@{}}%
\column{3}{@{}>{\hspre}l<{\hspost}@{}}%
\column{E}{@{}>{\hspre}l<{\hspost}@{}}%
\>[3]{}\Varid{r}\oplus_{M}(\Varid{ts}\oplus_{I}\Varid{s})\mathrel{=}(\Varid{r}\oplus_{M}\Varid{ts})\oplus_{I}\Varid{s}~~,{}\<[E]%
\ColumnHook
\end{hscode}\resethooks
where \ensuremath{\Varid{ts}} is in the range of \ensuremath{\Varid{dragonI}}. In particular, either \ensuremath{\Varid{ts}} is \ensuremath{[\mskip1.5mu \mskip1.5mu]} or it has an odd length.
With \ensuremath{(\Varid{s},\Varid{r})} being \ensuremath{(\Conid{C},\Conid{C})}, \ensuremath{(\Conid{C},\Conid{A})}, \ensuremath{(\Conid{A},\Conid{C})}, and \ensuremath{(\Conid{A},\Conid{A})},
the condition above expands respectively to:
\begingroup
\setlength{\mathindent}{0.5cm}
\begin{hscode}\SaveRestoreHook
\column{B}{@{}>{\hspre}l<{\hspost}@{}}%
\column{23}{@{}>{\hspre}l<{\hspost}@{}}%
\column{31}{@{}>{\hspre}l<{\hspost}@{}}%
\column{74}{@{}>{\hspre}l<{\hspost}@{}}%
\column{104}{@{}>{\hspre}l<{\hspost}@{}}%
\column{E}{@{}>{\hspre}l<{\hspost}@{}}%
\>[B]{}(\Varid{lr}\triangleright\Varid{ts}){}\<[23]%
\>[23]{}\mathbin{{+}\mskip-8mu{+}}[\mskip1.5mu \Conid{L}\mskip1.5mu]{}\<[31]%
\>[31]{}\mathbin{{+}\mskip-8mu{+}}\Varid{map}\;\Varid{inv}\;(\Varid{reverse}\;(\Varid{lr}\triangleright\Varid{ts})){}\<[74]%
\>[74]{}\mathrel{=}\Varid{lr}\triangleright(\Varid{ts}\mathbin{{+}\mskip-8mu{+}}[\mskip1.5mu \Conid{L}\mskip1.5mu]{}\<[104]%
\>[104]{}\mathbin{{+}\mskip-8mu{+}}\Varid{map}\;\Varid{inv}\;(\Varid{reverse}\;\Varid{ts}))~~,{}\<[E]%
\\
\>[B]{}(\Varid{rl}\triangleright\Varid{ts}){}\<[23]%
\>[23]{}\mathbin{{+}\mskip-8mu{+}}[\mskip1.5mu \Conid{L}\mskip1.5mu]{}\<[31]%
\>[31]{}\mathbin{{+}\mskip-8mu{+}}\Varid{map}\;\Varid{inv}\;(\Varid{reverse}\;(\Varid{rl}\triangleright\Varid{ts})){}\<[74]%
\>[74]{}\mathrel{=}\Varid{rl}\triangleright(\Varid{ts}\mathbin{{+}\mskip-8mu{+}}[\mskip1.5mu \Conid{L}\mskip1.5mu]{}\<[104]%
\>[104]{}\mathbin{{+}\mskip-8mu{+}}\Varid{map}\;\Varid{inv}\;(\Varid{reverse}\;\Varid{ts}))~~,{}\<[E]%
\\
\>[B]{}(\Varid{lr}\triangleright\Varid{ts}){}\<[23]%
\>[23]{}\mathbin{{+}\mskip-8mu{+}}[\mskip1.5mu \Conid{R}\mskip1.5mu]{}\<[31]%
\>[31]{}\mathbin{{+}\mskip-8mu{+}}\Varid{map}\;\Varid{inv}\;(\Varid{reverse}\;(\Varid{lr}\triangleright\Varid{ts})){}\<[74]%
\>[74]{}\mathrel{=}\Varid{lr}\triangleright(\Varid{ts}\mathbin{{+}\mskip-8mu{+}}[\mskip1.5mu \Conid{R}\mskip1.5mu]{}\<[104]%
\>[104]{}\mathbin{{+}\mskip-8mu{+}}\Varid{map}\;\Varid{inv}\;(\Varid{reverse}\;\Varid{ts}))~~,{}\<[E]%
\\
\>[B]{}(\Varid{rl}\triangleright\Varid{ts}){}\<[23]%
\>[23]{}\mathbin{{+}\mskip-8mu{+}}[\mskip1.5mu \Conid{R}\mskip1.5mu]{}\<[31]%
\>[31]{}\mathbin{{+}\mskip-8mu{+}}\Varid{map}\;\Varid{inv}\;(\Varid{reverse}\;(\Varid{rl}\triangleright\Varid{ts})){}\<[74]%
\>[74]{}\mathrel{=}\Varid{rl}\triangleright(\Varid{ts}\mathbin{{+}\mskip-8mu{+}}[\mskip1.5mu \Conid{R}\mskip1.5mu]{}\<[104]%
\>[104]{}\mathbin{{+}\mskip-8mu{+}}\Varid{map}\;\Varid{inv}\;(\Varid{reverse}\;\Varid{ts}))~~.{}\<[E]%
\ColumnHook
\end{hscode}\resethooks
\endgroup
All of them are instances of \eqref{eq:lr-interleave-distr}!
We have, in fact, proved the general case already.

\section{Conclusions and related work}

We have proved the equivalence of the two approaches to constructing the dragon curve.
Once the problem is properly formalised, the rest is a good exercise in letting the symbols do the work,
with distributing \ensuremath{(\Varid{lr}\,\triangleright)} inwards as the clear goal.
The two approaches to constructing generalised dragon curves are respectively a \ensuremath{\Varid{foldr}} and a \ensuremath{\Varid{foldl}}, and their equivalence is proved by the second duality theorem.

\citet{Tabachnikov:14:Dragon} gave a brief historical account of the dragon curve,
reviewed some of its properties, and proposed some open problems.
Many interesting properties of the dragon curve were presented by \citet{DavisKnuth:70:Number}.
While we have based our discussion on finite lists and proof by induction,
they discussed the dragon curve of infinite order:
\begin{hscode}\SaveRestoreHook
\column{B}{@{}>{\hspre}l<{\hspost}@{}}%
\column{E}{@{}>{\hspre}l<{\hspost}@{}}%
\>[B]{}\Varid{dragon}\mathbin{::}\Conid{Stream}\;\Conid{Turn}{}\<[E]%
\\
\>[B]{}\Varid{dragon}\mathrel{=}\Varid{lr}\triangleright\Varid{dragon}~~,{}\<[E]%
\ColumnHook
\end{hscode}\resethooks
of which a dragon curve of any finite order is a prefix.
It will be interesting to see whether proof techniques for codata, for example \citet{Hinze:09:Reasoning},
can be applied to prove some of the properties in \citet{DavisKnuth:70:Number}.

Many fractal curves can be constructed in two ways.
For instance, the Koch snowflake can be constructed by either making four copies of the previous shape and joining them together, or by splitting one segment into four segments that form a spike.
The former approach is often described by a recursive function involving copying and rotation,
while the latter by a \citeauthor{Lindenmayer:68:Mathematical} system \citeyearpar{Lindenmayer:68:Mathematical}.
It remains to be explored whether we can develop a theory relating the two styles of construction for such fractals in general.

\paragraph{Acknowledgements}
The third author discussed initial ideas of this pearl with Robby Findler,
who offered plenty of suggestions.
The reviewers carefully examined and redid our proofs (two of them produced their
own Agda formalisations), suggested cleaner approaches,
pointed out details which we should not sweep under the rug,
and suggested a new, catchy title for this pearl.
The Agda formalisation we supply with the pearl is adapted from one developed by a reviewer,
who kindly allowed us to do so.
To them we owe our sincere gratitude.

\bibliographystyle{ACM-Reference-Format}
\bibliography{dragon.bib}

\appendix

\section{Proofs of distributive properties}
\label{sec:proof-distributivity}

For completeness we present the proofs of properties \eqref{eq:interleave-cat} and \eqref{eq:interleave-reverse}.

\vspace{1em}
\noindent \emph{Proof of \eqref{eq:interleave-cat}}:
\ensuremath{\Varid{alts}\;\Varid{t}\triangleright(\Varid{xs}\mathbin{{+}\mskip-8mu{+}}[\mskip1.5mu \Varid{z}\mskip1.5mu]\mathbin{{+}\mskip-8mu{+}}\Varid{ys})\mathrel{=}(\Varid{alts}\;\Varid{t}\triangleright\Varid{xs})\mathbin{{+}\mskip-8mu{+}}[\mskip1.5mu \Varid{z}\mskip1.5mu]\mathbin{{+}\mskip-8mu{+}}(\Varid{alts}\;\Varid{t}\triangleright\Varid{ys})}, for odd-length \ensuremath{\Varid{xs}}.
Induction on \ensuremath{\Varid{xs}}.

\noindent{\bf Case} \ensuremath{\Varid{xs}\mathbin{:=}[\mskip1.5mu \Varid{x}\mskip1.5mu]}
\begin{hscode}\SaveRestoreHook
\column{B}{@{}>{\hspre}l<{\hspost}@{}}%
\column{3}{@{}>{\hspre}c<{\hspost}@{}}%
\column{3E}{@{}l@{}}%
\column{6}{@{}>{\hspre}l<{\hspost}@{}}%
\column{10}{@{}>{\hspre}l<{\hspost}@{}}%
\column{E}{@{}>{\hspre}l<{\hspost}@{}}%
\>[6]{}\Varid{alts}\;\Varid{t}\triangleright(\Varid{x}\mathbin{:}\Varid{z}\mathbin{:}\Varid{ys}){}\<[E]%
\\
\>[3]{}\mathrel{=}{}\<[3E]%
\>[10]{}\mbox{\commentbegin  definitions of \ensuremath{\Varid{alts}} and \ensuremath{(\triangleright)}, \ensuremath{\Varid{inv}\;(\Varid{inv}\;\Varid{t})\mathrel{=}\Varid{t}}  \commentend}{}\<[E]%
\\
\>[3]{}\hsindent{3}{}\<[6]%
\>[6]{}\Varid{t}\mathbin{:}\Varid{x}\mathbin{:}\Varid{inv}\;\Varid{t}\mathbin{:}\Varid{z}\mathbin{:}(\Varid{alts}\;\Varid{t}\triangleright\Varid{ys}){}\<[E]%
\\
\>[3]{}\mathrel{=}{}\<[3E]%
\>[10]{}\mbox{\commentbegin  definitions of \ensuremath{\Varid{alts}}, \ensuremath{(\mathbin{{+}\mskip-8mu{+}})}, and \ensuremath{(\triangleright)}  \commentend}{}\<[E]%
\\
\>[3]{}\hsindent{3}{}\<[6]%
\>[6]{}(\Varid{alts}\;\Varid{t}\triangleright[\mskip1.5mu \Varid{x}\mskip1.5mu])\mathbin{{+}\mskip-8mu{+}}[\mskip1.5mu \Varid{z}\mskip1.5mu]\mathbin{{+}\mskip-8mu{+}}(\Varid{alts}\;\Varid{t}\triangleright\Varid{ys})~~.{}\<[E]%
\ColumnHook
\end{hscode}\resethooks
\noindent{\bf Case} \ensuremath{\Varid{xs}\mathbin{:=}\Varid{x}_{0}\mathbin{:}\Varid{x}_{1}\mathbin{:}\Varid{xs}}
\begin{hscode}\SaveRestoreHook
\column{B}{@{}>{\hspre}l<{\hspost}@{}}%
\column{3}{@{}>{\hspre}c<{\hspost}@{}}%
\column{3E}{@{}l@{}}%
\column{6}{@{}>{\hspre}l<{\hspost}@{}}%
\column{9}{@{}>{\hspre}l<{\hspost}@{}}%
\column{E}{@{}>{\hspre}l<{\hspost}@{}}%
\>[6]{}\Varid{alts}\;\Varid{t}\triangleright(\Varid{x}_{0}\mathbin{:}\Varid{x}_{1}\mathbin{:}\Varid{xs}\mathbin{{+}\mskip-8mu{+}}[\mskip1.5mu \Varid{z}\mskip1.5mu]\mathbin{{+}\mskip-8mu{+}}\Varid{ys}){}\<[E]%
\\
\>[3]{}\mathrel{=}{}\<[3E]%
\>[9]{}\mbox{\commentbegin  definition of \ensuremath{\Varid{alts}} and \ensuremath{(\triangleright)}, \ensuremath{\Varid{inv}\;(\Varid{inv}\;\Varid{t})\mathrel{=}\Varid{t}}  \commentend}{}\<[E]%
\\
\>[3]{}\hsindent{3}{}\<[6]%
\>[6]{}\Varid{t}\mathbin{:}\Varid{x}_{0}\mathbin{:}\Varid{inv}\;\Varid{t}\mathbin{:}\Varid{x}_{1}\mathbin{:}(\Varid{alts}\;\Varid{t}\triangleright(\Varid{xs}\mathbin{{+}\mskip-8mu{+}}[\mskip1.5mu \Varid{z}\mskip1.5mu]\mathbin{{+}\mskip-8mu{+}}\Varid{ys})){}\<[E]%
\\
\>[3]{}\mathrel{=}{}\<[3E]%
\>[9]{}\mbox{\commentbegin  induction hypothesis  \commentend}{}\<[E]%
\\
\>[3]{}\hsindent{3}{}\<[6]%
\>[6]{}\Varid{t}\mathbin{:}\Varid{x}_{0}\mathbin{:}\Varid{inv}\;\Varid{t}\mathbin{:}\Varid{x}_{1}\mathbin{:}(\Varid{alts}\;\Varid{t}\triangleright\Varid{xs})\mathbin{{+}\mskip-8mu{+}}[\mskip1.5mu \Varid{z}\mskip1.5mu]\mathbin{{+}\mskip-8mu{+}}(\Varid{alts}\;\Varid{t}\triangleright\Varid{ys}){}\<[E]%
\\
\>[3]{}\mathrel{=}{}\<[3E]%
\>[9]{}\mbox{\commentbegin  definitions of \ensuremath{\Varid{alts}}, \ensuremath{(\mathbin{{+}\mskip-8mu{+}})}, and \ensuremath{(\triangleright)}, \ensuremath{\Varid{inv}\;(\Varid{inv}\;\Varid{t})\mathrel{=}\Varid{t}}  \commentend}{}\<[E]%
\\
\>[3]{}\hsindent{3}{}\<[6]%
\>[6]{}(\Varid{alts}\;\Varid{t}\triangleright(\Varid{x}_{0}\mathbin{:}\Varid{x}_{1}\mathbin{:}\Varid{xs}))\mathbin{{+}\mskip-8mu{+}}[\mskip1.5mu \Varid{z}\mskip1.5mu]\mathbin{{+}\mskip-8mu{+}}(\Varid{alts}\;\Varid{t}\triangleright\Varid{ys})~~.{}\<[E]%
\ColumnHook
\end{hscode}\resethooks

In our Agda proof, we define a predicate \ensuremath{\Conid{OddLen}} below:
\begin{hscode}\SaveRestoreHook
\column{B}{@{}>{\hspre}l<{\hspost}@{}}%
\column{5}{@{}>{\hspre}l<{\hspost}@{}}%
\column{13}{@{}>{\hspre}l<{\hspost}@{}}%
\column{32}{@{}>{\hspre}l<{\hspost}@{}}%
\column{34}{@{}>{\hspre}l<{\hspost}@{}}%
\column{E}{@{}>{\hspre}l<{\hspost}@{}}%
\>[B]{}\mathbf{data}\;\Conid{OddLen}\;\{\mskip1.5mu \Conid{A}\mathbin{:}\Conid{Set}\mskip1.5mu\}\mathbin{:}\Conid{List}\;\Conid{A}\to \Conid{Set}\;\mathbf{where}{}\<[E]%
\\
\>[B]{}\hsindent{5}{}\<[5]%
\>[5]{}\lbrack\_\rbrack_{o}{}\<[13]%
\>[13]{}\mathbin{:}\forall\;\Varid{x}{}\<[32]%
\>[32]{}\to \Conid{OddLen}\;(\Varid{x}\mathbin{::}[\mskip1.5mu \mskip1.5mu]){}\<[E]%
\\
\>[B]{}\hsindent{5}{}\<[5]%
\>[5]{}\_\!::_{o}\!\_::\!\_{}\<[13]%
\>[13]{}\mathbin{:}\forall\;\Varid{x}_{0}\;\Varid{x}_{1}\;\{\mskip1.5mu \Varid{xs}\mskip1.5mu\}{}\<[34]%
\>[34]{}\to \Conid{OddLen}\;\Varid{xs}\to \Conid{OddLen}\;(\Varid{x}_{0}\mathbin{::}\Varid{x}_{1}\mathbin{::}\Varid{xs})~~.{}\<[E]%
\ColumnHook
\end{hscode}\resethooks
and perform induction on \ensuremath{\Conid{OddLen}\;\Varid{xs}}; thus, the base case is a singleton list while the inductive case matches the pattern \ensuremath{\Varid{x}_{0}\mathbin{::}\Varid{x}_{1}\mathbin{::}\Varid{xs}}.

\vspace{2em}
\noindent\emph{Proof of \eqref{eq:interleave-reverse}}: \ensuremath{\Varid{reverse}\;(\Varid{alts}\;\Varid{t}\triangleright\Varid{xs})\mathrel{=}\Varid{alts}\;(\Varid{inv}\;\Varid{t})\triangleright\Varid{reverse}\;\Varid{xs}}, for odd-length \ensuremath{\Varid{xs}}.
Induction on \ensuremath{\Varid{xs}}.
We omit the base case.
For the inductive case \ensuremath{\Varid{xs}\mathbin{:=}\Varid{x}_{0}\mathbin{:}\Varid{x}_{1}\mathbin{:}\Varid{xs}}:
\begin{hscode}\SaveRestoreHook
\column{B}{@{}>{\hspre}l<{\hspost}@{}}%
\column{7}{@{}>{\hspre}l<{\hspost}@{}}%
\column{9}{@{}>{\hspre}l<{\hspost}@{}}%
\column{E}{@{}>{\hspre}l<{\hspost}@{}}%
\>[7]{}\Varid{reverse}\;(\Varid{alts}\;\Varid{t}\triangleright(\Varid{x}_{0}\mathbin{:}\Varid{x}_{1}\mathbin{:}\Varid{xs})){}\<[E]%
\\
\>[B]{}\mathrel{=}{}\<[9]%
\>[9]{}\mbox{\commentbegin  definitions of \ensuremath{\Varid{alts}} and \ensuremath{(\triangleright)}, \ensuremath{\Varid{inv}\;(\Varid{inv}\;\Varid{t})\mathrel{=}\Varid{t}}  \commentend}{}\<[E]%
\\
\>[B]{}\hsindent{7}{}\<[7]%
\>[7]{}\Varid{reverse}\;(\Varid{t}\mathbin{:}\Varid{x}_{0}\mathbin{:}\Varid{inv}\;\Varid{t}\mathbin{:}\Varid{x}_{1}\mathbin{:}(\Varid{alts}\;\Varid{t}\triangleright\Varid{xs})){}\<[E]%
\\
\>[B]{}\mathrel{=}{}\<[9]%
\>[9]{}\mbox{\commentbegin  definition of \ensuremath{\Varid{reverse}}  \commentend}{}\<[E]%
\\
\>[B]{}\hsindent{7}{}\<[7]%
\>[7]{}\Varid{reverse}\;(\Varid{alts}\;\Varid{t}\triangleright\Varid{xs})\mathbin{{+}\mskip-8mu{+}}[\mskip1.5mu \Varid{x}_{1},\Varid{inv}\;\Varid{t},\Varid{x}_{0},\Varid{t}\mskip1.5mu]{}\<[E]%
\\
\>[B]{}\mathrel{=}{}\<[9]%
\>[9]{}\mbox{\commentbegin  induction hypothesis  \commentend}{}\<[E]%
\\
\>[B]{}\hsindent{7}{}\<[7]%
\>[7]{}(\Varid{alts}\;(\Varid{inv}\;\Varid{t})\triangleright\Varid{reverse}\;\Varid{xs})\mathbin{{+}\mskip-8mu{+}}[\mskip1.5mu \Varid{x}_{1},\Varid{inv}\;\Varid{t},\Varid{x}_{0},\Varid{t}\mskip1.5mu]{}\<[E]%
\\
\>[B]{}\mathrel{=}{}\<[9]%
\>[9]{}\mbox{\commentbegin  definition of \ensuremath{(\triangleright)}  \commentend}{}\<[E]%
\\
\>[B]{}\hsindent{7}{}\<[7]%
\>[7]{}(\Varid{alts}\;(\Varid{inv}\;\Varid{t})\triangleright\Varid{reverse}\;\Varid{xs})\mathbin{{+}\mskip-8mu{+}}[\mskip1.5mu \Varid{x}_{1}\mskip1.5mu]\mathbin{{+}\mskip-8mu{+}}(\Varid{alts}\;(\Varid{inv}\;\Varid{t})\triangleright[\mskip1.5mu \Varid{x}_{0}\mskip1.5mu]){}\<[E]%
\\
\>[B]{}\mathrel{=}{}\<[9]%
\>[9]{}\mbox{\commentbegin  since \ensuremath{\Varid{reverse}} preserves length, by \eqref{eq:interleave-cat}  \commentend}{}\<[E]%
\\
\>[B]{}\hsindent{7}{}\<[7]%
\>[7]{}\Varid{alts}\;(\Varid{inv}\;\Varid{t})\triangleright(\Varid{reverse}\;\Varid{xs}\mathbin{{+}\mskip-8mu{+}}[\mskip1.5mu \Varid{x}_{1},\Varid{x}_{0}\mskip1.5mu]){}\<[E]%
\\
\>[B]{}\mathrel{=}{}\<[9]%
\>[9]{}\mbox{\commentbegin  definition of \ensuremath{\Varid{reverse}}   \commentend}{}\<[E]%
\\
\>[B]{}\hsindent{7}{}\<[7]%
\>[7]{}\Varid{alts}\;(\Varid{inv}\;\Varid{t})\triangleright\Varid{reverse}\;(\Varid{x}_{0}\mathbin{:}\Varid{x}_{1}\mathbin{:}\Varid{xs})~~.{}\<[E]%
\ColumnHook
\end{hscode}\resethooks
\end{document}